\documentclass[11pt]{article}

\pdfoutput=1
\usepackage{graphicx}
\usepackage{hyperref}
\usepackage{geometry}
\usepackage{color}
\usepackage{caption}
\usepackage{latexsym}
\usepackage{amssymb}
\usepackage{epsf}
\usepackage{amsmath}
\usepackage{graphicx}
\usepackage{slashed}
\usepackage{float}
\usepackage{cases}
\usepackage[square,comma,numbers,sort&compress]{natbib}
\begin{document}

\title{\bf $ttH$ Anomalous Coupling in Double Higgs Production}
\author{ {Kenji Nishiwaki,\footnote{e-mail: nishiwaki@hri.res.in} \ 
          Saurabh Niyogi\,\footnote{e-mail: sourabh@hri.res.in} \ 
          and Ambresh Shivaji\,\footnote{e-mail: ambreshkshivaji@hri.res.in} } \\[6pt]
          {Regional Centre for Accelerator-based Particle Physics} \\
         {\it Harish-Chandra Research Institute} \\ 
         {\it Chhatnag Road, Junsi, Allahabad-211019, India}
        }
\maketitle{}

 \begin{abstract}
 \noindent
  We study the effects of top-Higgs anomalous coupling in the production of a pair
  of Higgs boson via gluon fusion at the Large Hadron Collider (LHC). The introduction of 
  anomalous $ttH$ coupling can alter the hadronic double Higgs boson cross section and  
  can lead to characteristic changes in certain kinematic distributions. 
  We perform a global analysis based on available LHC data on the Higgs to constrain the parameters
  of $ttH$ anomalous coupling. Possible overlap of the predictions due to anomalous $ttH$ coupling 
  with those due to anomalous trilinear Higgs coupling is also studied. 
  We briefly discuss the effect of the anomalous $ttH$ coupling on the $HZ$ production 
  via gluon fusion which is one of the main backgrounds in the $HH \to \gamma\gamma b {\bar b}$ channel.
  
 \end{abstract}

\vfill
\begin{flushright}
HRI-P-13-09-002\\
RECAPP-HRI-2013-020
\end{flushright}

\newpage

\section{Introduction}

The discovery of Higgs boson by the Large Hadron Collider (LHC) \cite{Aad:2012tfa,Chatrchyan:2012ufa} at CERN once again writes
the great success story of the standard model (SM). Though it is not yet 
conclusively declared that this is the `very' Higgs boson postulated in the standard model, 
but more data consolidate the same. However, there are still room left
 for new physics to show up at the weak scale within the reach of the LHC. 
There are many ways to search for new physics at the LHC. The most popular one is
to look for new resonances directly produced in proton-proton collision. But
no such new particles have been found till date. Hence, lower limits at 95\% confidence level (CL) has been placed
constraining various models of new physics. The other way is to look for deviation in couplings where new physics
effects may enter. It will, in turn, show up in appropriate production or decay processes at the LHC.
We shall take this latter approach in a model independent way to probe the nature of new physics. 
In fact, after the discovery of the Higgs boson, it still remains to verify its couplings 
with other standard model particles and also with itself. Recent studies involving
anomalous couplings of the Higgs boson at the LHC have been reported in \cite{recent_globalanalysis}.
 Prospects of the measurement potential of various Higgs couplings at future linear collider are also discussed in 
 Ref.~\cite{Klute:2013cx}.

In the standard model, the couplings of the Higgs boson with the fermions and gauge bosons 
are proportional to their masses. Its large $\sim O(1)$ coupling with top quark 
is the reason for expecting that any deviation, if present, might show up via top-Higgs
coupling. Hence, probing this coupling always remains a priority. 
{The top-Higgs Yukawa coupling can be indirectly probed by the measurements of inclusive
Higgs boson production which is dominated by gluon fusion process and also in the decay of the Higgs
to diphoton and digluon channels mediated by the top quark loop. However, the only direct way to constrain this coupling is to 
measure $ttH$ production at the LHC.
ATLAS and CMS has already published data in this direction, but
not much deviation from the standard model has been observed. Given the theoretical and experimental uncertainties, 
it is difficult to derive any meaningful limit from the collected data~\cite{Klute:2012pu,Adelman:2013gis}.

Due to the presence of new physics the top-Higgs coupling can differ from its standard model 
value \cite{Feng:2003uv,Frederix:2007gi,Gabrielli:2010cw,Dawson:2012mk}. 
These deviations can come from
higher dimensional operators present below a certain scale \cite{Pierce:2006dh,AguilarSaavedra:2009mx,Adelman:2013gis,Degrande:2012gr}. 
Moreover, many of the new physics models also predict deviation in $ttH$ coupling from the standard one.}
{The standard model Higgs boson is predicted to be CP-even. However, LHC data do not rule out the Higgs to be a mixed 
CP state. Taking this freedom, we consider top-Higgs coupling to  be CP-violating one for this work.
We stress that} we do not focus on some specific model or some set of effective operators. 
Instead, we consider a general parameterization of anomalous top-Higgs 
coupling which definitely includes all the above effects.

Double Higgs production at the LHC provides a good opportunity to probe various couplings of the Higgs boson. 
Since gluon fusion is still the dominant channel for Higgs pair production, just like single Higgs production, this
process has strong dependence on $ttH$ coupling. At the same time, it can give access to the Higgs trilinear 
coupling as well.

This paper is organized as follows.
In the following section~\ref{Section:2}, we discuss the Higgs pair production in the standard model itself.
In section~\ref{Section:3}, the general parameterization of top-Higgs interaction is motivated.
This will be followed by the effects of anomalous coupling on the production cross section and on 
different kinematic variables of the Higgs pair production at the LHC. 
Next, section~\ref{Section:5} will consist of the constraints from the LHC experiments and resultant global analysis.
Finally in section~\ref{Section:6}, we summarize our observations and give careful consideration to the prospects of 
the Higgs pair production based on the results of the global analysis.

\section{Higgs pair production in the standard model
\label{Section:2}}

\begin{figure}[t]
\begin{center}
\includegraphics[width=10cm]{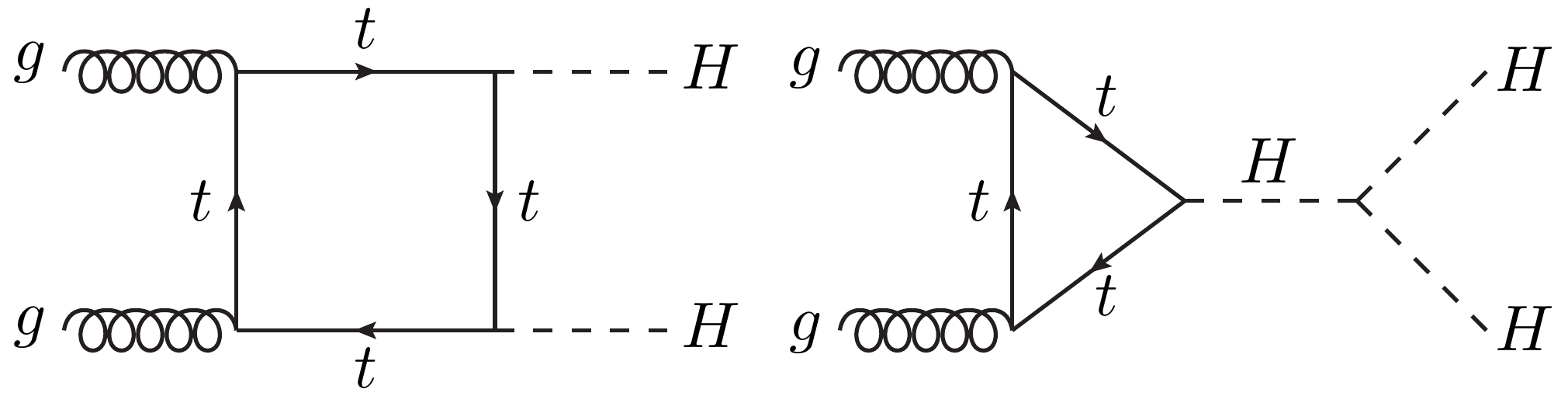}
\caption{ {Prototype diagrams for the leading order production of double Higgs via gluon fusion. Other diagrams are generated by permuting the external legs appropriately.}}
\label{fig:gghh}
\end{center}
\end{figure}

\begin{figure}[t]
\centering
\begin{minipage}{.5\textwidth}
  \centering
  \includegraphics[width=7.5cm]{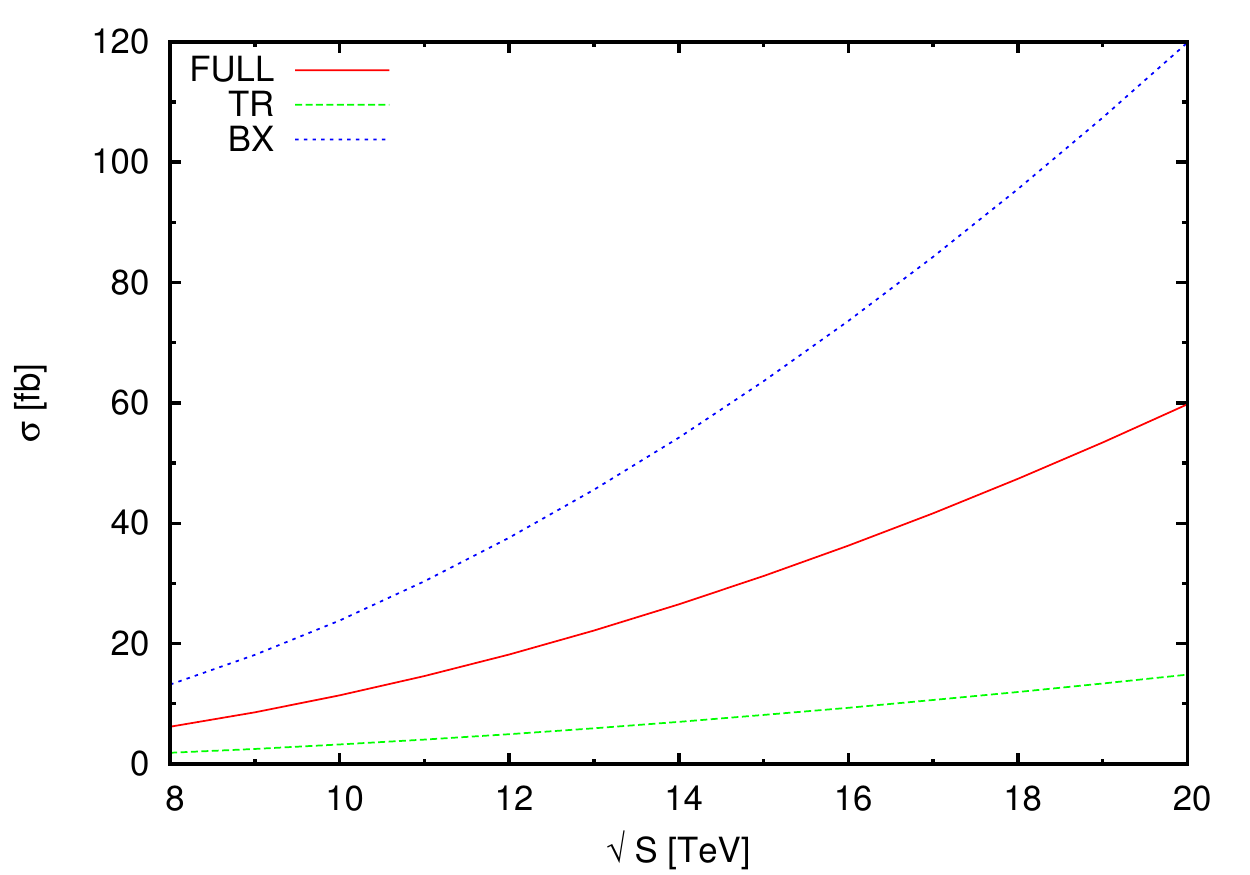}
  \captionsetup{width=.8\textwidth}
   \caption{Triangle (TR) and box (BX) amplitudes contributions to the hadronic cross sections at various collider center-of-mass energies in the standard model.}
  \label{fig:cmeplotsm}
\end{minipage}%
\begin{minipage}{.5\textwidth}
  \centering
  \includegraphics[width=7.5cm]{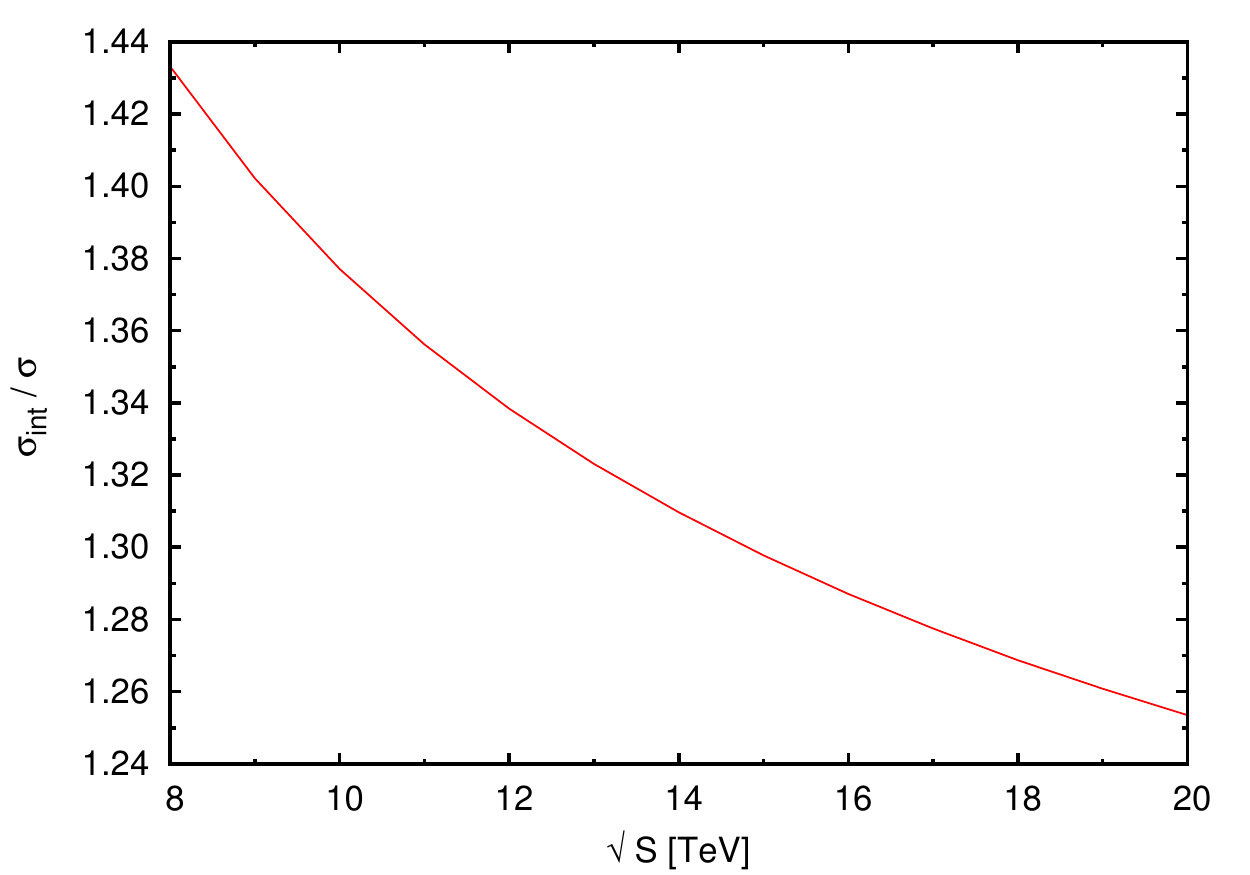}
  \captionsetup{width=.8\textwidth}
  \caption{Fractional contribution of the cross section due to the interference term 
           at various collider center-of-mass energies in the standard model.}
  \label{fig:ratioplotsm}
\end{minipage}
\end{figure}

The Higgs boson pair production within the standard model was first 
studied in~\cite{Glover:1987nx,Eboli:1987dy}.
Very much like the production of single Higgs boson, the gluon fusion
channel is the dominant mode to produce a pair of Higgs boson at hadron colliders.
At the leading order the process proceeds via quark loop diagrams, shown
in Fig.~\ref{fig:gghh}.\footnote{
These diagrams are drawn using the Jaxodraw package~\cite{Binosi:2008ig}.} 
The major contribution to the hadronic 
cross section comes from the top quark loop diagram. The 
bottom quark loop contribution is well below $1 \%$ ($0.2 \%$ at 14 TeV) of the total 
cross section.
One of the important features of this
process is the destructive interference that takes places between the box and 
the triangle contributions.\footnote{
We use, $\sigma = \sigma_{tr} + \sigma_{bx} - \sigma_{int}$, where $\sigma_{int}$ 
is due to the interference between the triangle and box amplitudes.} 
The two contributions are separately gauge invariant.
As we can see in Fig.~\ref{fig:cmeplotsm}, the destructive interference effect is quite strong. 
For example, at 14 TeV, the separate 
contributions of the triangle and box amplitudes towards the total hadronic cross 
section is about $6.98$ fb and $54.22$ fb respectively. The net cross section, on the other hand, 
is only 26.50 fb, {\it i.e.}, there is a 
reduction of more than $50 \%$ in the cross section due to the interference term.  
Note that the minimum threshold to produce the Higgs boson pair is greater than the Higgs mass, therefore, 
the intermediate Higgs boson in the triangle diagram is always off-shell.
We expect that due to the propagator suppression in the triangle amplitude, the interference effect falls 
at higher energies, see Fig.~\ref{fig:ratioplotsm}.

Higgs pair production has also been a subject of discussion in the context of various new physics models~\cite{Kanemura:2008ub,Contino:2012xk,Goertz:2013kp,Barr:2013tda}
including the minimal supersymmetric standard model (MSSM)~\cite{Plehn:1996wb} and the Little Higgs~\cite{little_higgs}.
{Total Higgs pair production cross section including higher order corrections
has been discussed in~\cite{Dawson:1998py,Baglio:2012np,Shao:2013bz,Grigo:2013rya,deFlorian:2013jea}.}
It is known that in the large fermion mass limit the amplitude does not vanish. This non-decoupling behaviour 
makes the process sensitive to the existence of heavier quarks in new physics models~\cite{Asakawa:2010xj}.
The process is also important from the point of view of measuring the trilinear self-coupling of 
the Higgs boson~\cite{Djouadi:1999ei} which is present in the triangle diagram of Fig.~\ref{fig:gghh}. The precise 
measurement of the trilinear self-coupling of the Higgs boson is required to confirm the form 
of the scalar potential responsible for the electroweak symmetry breaking.
However, the collider center-of-mass energy and the luminosity required to observe 
this channel at the LHC has not been reached yet. 

\section{The top-Higgs anomalous coupling
\label{Section:3}}
It is well known that the absolute sign of the standard model Yukawa coupling is arbitrary. 
Nevertheless, its relative sign with respect to the mass term is completely determined. 
Any change in this relative sign will be a clear indication of new physics effects. At the same time, 
this change in the relative sign can have serious implications for those processes which involve both $ttH$ and 
any of the three couplings $HWW$, $HZZ$ and $HHH$. 
Plausibility of such scenarios has been considered in associated production of a single top and a 
Higgs boson at the LHC~\cite{Barger:2009ky,Biswas:2012bd,Farina:2012xp,Agrawal:2012ga}.
Since, the Higgs pair production process involves both the top-Yukawa coupling and the trilinear 
Higgs coupling, the relative sign change between the two couplings will lead to constructive
interference between the box and the triangle contributions. As a result the Higgs pair production rates 
at the LHC will be higher as compared to those predicted in the standard model. In addition to that, 
the presence of new physics can also modify the nature of various standard model couplings. 
The top-quark being exceptionally heavy as compared to the other fermions may hold the
signatures of new physics. 
In the standard model, the top-Yukawa coupling is purely scalar type. 
Many new physics models, such as the composite Higgs models~\cite{Koulovassilopoulos:1993pw} and models with the extended Higgs 
sector~\cite{Accomando:2006ga} suggest that the Yukawa couplings can be an admixture of both the scalar and 
pseudoscalar type of couplings. In other words, the physical Higgs boson may not have a definite 
CP property~\cite{Bernreuther:1998qv}. 


A phenomenological Lagrangian describing the nonstandard top quark Yukawa coupling 
can be parameterized as, 
\begin{equation}\label{eq:tth_anml}
 {\cal L}_{\rm ttH} = -\frac{g_w m_t}{2 M_w} \;\; {\bar t} (a + i b \gamma^5) t \; H,
\end{equation}
where $g_w$ is the $SU(2)$ gauge coupling constant. Both the dimensionless parameters $a$ and $b$ are real  
and they assume values 1 and 0 respectively in the standard model at the leading order. The 
$\gamma^5$ or the pseudoscalar part of the coupling has to be imaginary due to the hermiticity of the Lagrangian.
Since, CP is not an exact symmetry of the standard model, the CP-odd term, in principle, can be 
generated at higher loops. However, such contributions are expected to be very small within the
standard model. The above form of the top-Higgs coupling can also be motivated in the 
effective Lagrangian approach to new physics studies. In this approach the new physics effects 
can be parameterized by a set of gauge invariant higher dimensional operators involving the standard
model fields only. We can write down an effective Lagrangian using these operators 
as, 
\begin{equation}\label{eq:Leff}
 {\cal L_{\rm eff}} = \sum_i \frac{C_i}{\Lambda ^{d_i-4}} {\cal O}^i,
\end{equation}
where $d_i > 4$ is the mass dimension of the operator ${\cal O}^i$, the free parameter $C_i$ 
fixes the strength of the corresponding operator and $\Lambda$ is the cutoff scale above which this 
effective description of new physics is not valid. These higher dimensional operators can modify both the 
strength and the nature of various standard model couplings. For example, the {\it lowest} higher
dimensional operators which contribute to the top-Higgs Yukawa coupling are dimension-six operators
~\cite{AguilarSaavedra:2009mx,Degrande:2012gr,Adelman:2013gis} and these are given by
\begin{equation}\label{eq:6DOP}
 (\Phi^{\dagger}\Phi)( \bar Q_{L} t_{R} \tilde \Phi)\; ; \;  (\Phi^{\dagger}\sigma^I D_\mu \Phi)( \bar Q_{L} \gamma^\mu \sigma^I Q_{L})\; ; \;
 (\Phi^{\dagger} D_\mu \Phi)( \bar Q_{L} \gamma^\mu Q_{L}) \; ;\;  (\Phi^{\dagger} D_\mu \Phi)( \bar t_{R} \gamma^\mu t_{R}). 
\end{equation}
In the above, $\Phi\; (\tilde \Phi$ = $i \sigma_{2} \Phi^*)$ is the standard model Higgs doublet field, 
${\bar Q_L} = ({\bar t_L}, {\bar b_L})$ is the third generation quark doublet, $t_R$ is the top quark
singlet and  $\sigma^I (I=1,2,3)$ are the $2\times 2$ Pauli matrices. As a result of the electroweak symmetry breakdown, the field 
$\Phi$ obtains a vacuum expectation value and the above operators effectively generate deviations in the parameters of  
 Eq.~(\ref{eq:tth_anml}) away from their standard model values. We can assume similar parameterization for 
 other Yukawa couplings also. However, for our process under consideration, it is the top-Yukawa coupling which is the most relevant.

At present, there are no significant {\it direct} bounds on the anomalous top-Higgs coupling parameters 
from the collider experiments. In Ref.~\cite{Whisnant:1994fh} unitarity constraints on these parameters 
are derived assuming the new physics scale at 1 TeV which allow $O(1)$ values for the parameters.
Note that the parametric form of the anomalous $ttH$ coupling in Eq.~\ref{eq:tth_anml} violates 
the CP symmetry explicitly for non-zero $b$. The CP-odd part of the coupling contributes to both the 
electroweak baryogenesis and the electric dipole moments (EDMs) of fermions~\cite{Shu:2013uua,Zhang:1994fb,Brod:2013cka}. We can use the measurements 
of the EDMs of the electron and the neutron to place indirect bounds on the parameter $b$. 
In Ref.~\cite{Brod:2013cka}, the EDM bounds on $b$ are found to be of $O$(0.01). This bound can be 
circumvented if the electron, up and down quark Yukawa couplings are also anomalous. The phenomenology of 
top-Higgs anomalous coupling under consideration has been studied at both the linear~\cite{Zhang:1994fb,Godbole:2011hw} 
and hadron colliders~\cite{Schmidt:1992et,Ellis:2013yxa}. 
Now we consider the effect of top-Higgs anomalous coupling  on 
the Higgs pair production process, keeping all the other standard model couplings intact. However, 
in section 6, we will briefly discuss the effect of anomalous trilinear Higgs coupling in the same 
process.

\section{Higgs pair production in presence of anomalous $ttH$ coupling
\label{Section:4}}

The full amplitude of our process in presence of the anomalous $ttH$ coupling 
can be expressed in the following form,
\begin{equation}\label{M_ggHH}
 {\cal M} = a^2 {\cal M}_{bx}^{\rm SM} + b^2 {\cal M}_{bx}^{(1)} + ab {\cal M}_{bx}^{(2)} 
            + a {\cal M}_{tr}^{\rm SM} + b {\cal M}_{tr}^{(3)}.
\end{equation}
We consider this structure of the amplitude after computing the quark loop traces of the 
diagrams. Here,  
${\cal M}_{bx/tr}^{\rm SM}$ are the standard model values of the box ({\it bx}) and triangle ({\it tr}) amplitudes and 
${\cal M}_{bx/tr}^{\rm (i)}$ are the additional box and triangle contributions due to the 
pseudoscalar coupling of the Higgs boson with the top quark.
The terms linear in $b$ in the above amplitude are proportional to possible $\epsilon$-tensor 
structures such as $\epsilon(p_i,p_j,e_1,e_2)$ and $\epsilon(p_1,p_2,p_3,e_i)$, where $e_i$s are 
the polarization vectors of the gluons.\footnote{
$\epsilon(p_1,p_2,e_1,e_2) = \epsilon^{\mu\nu\alpha\beta}p_{1\mu}p_{2\nu}e_{1\alpha}e_{2\beta}.$} 
The amplitude-squared will also have terms odd in $b$. However, once the gluon 
polarizations are summed over, such terms in the amplitude-squared vanish due to the 4-momentum 
conservation. Thus the unpolarized cross section of the two Higgs production process is expected 
to depend only on the absolute value of the parameter $b$. On the other hand, a change in sign in the 
parameter $a$ leads to significant changes in results discussed below. 
   \begin{figure}[t]
\centering
\begin{minipage}{.5\textwidth}
  \centering
  \includegraphics[width=7.5cm]{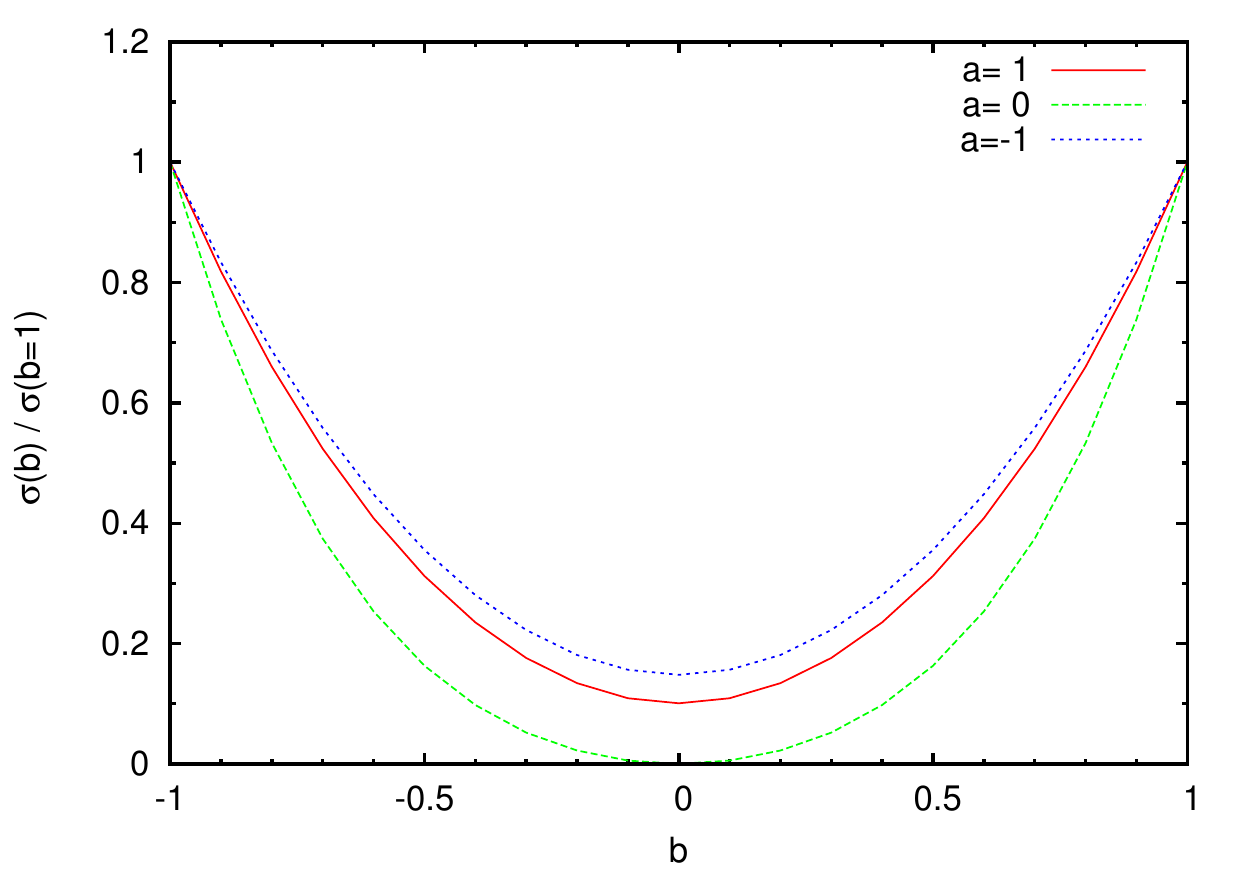}
  \captionsetup{width=.8\textwidth}
   \caption{Cross sections as function of parameter $b$ for $a=1,0,-1$. We have scaled the cross sections in all the 
   three cases by their maximum values at $b=1$. The symmetry of these plots about $b=0$ is explained in the text.}
  \label{fig:scanplothh}
\end{minipage}%
\begin{minipage}{.5\textwidth}
  \centering
  \includegraphics[width=7.5cm]{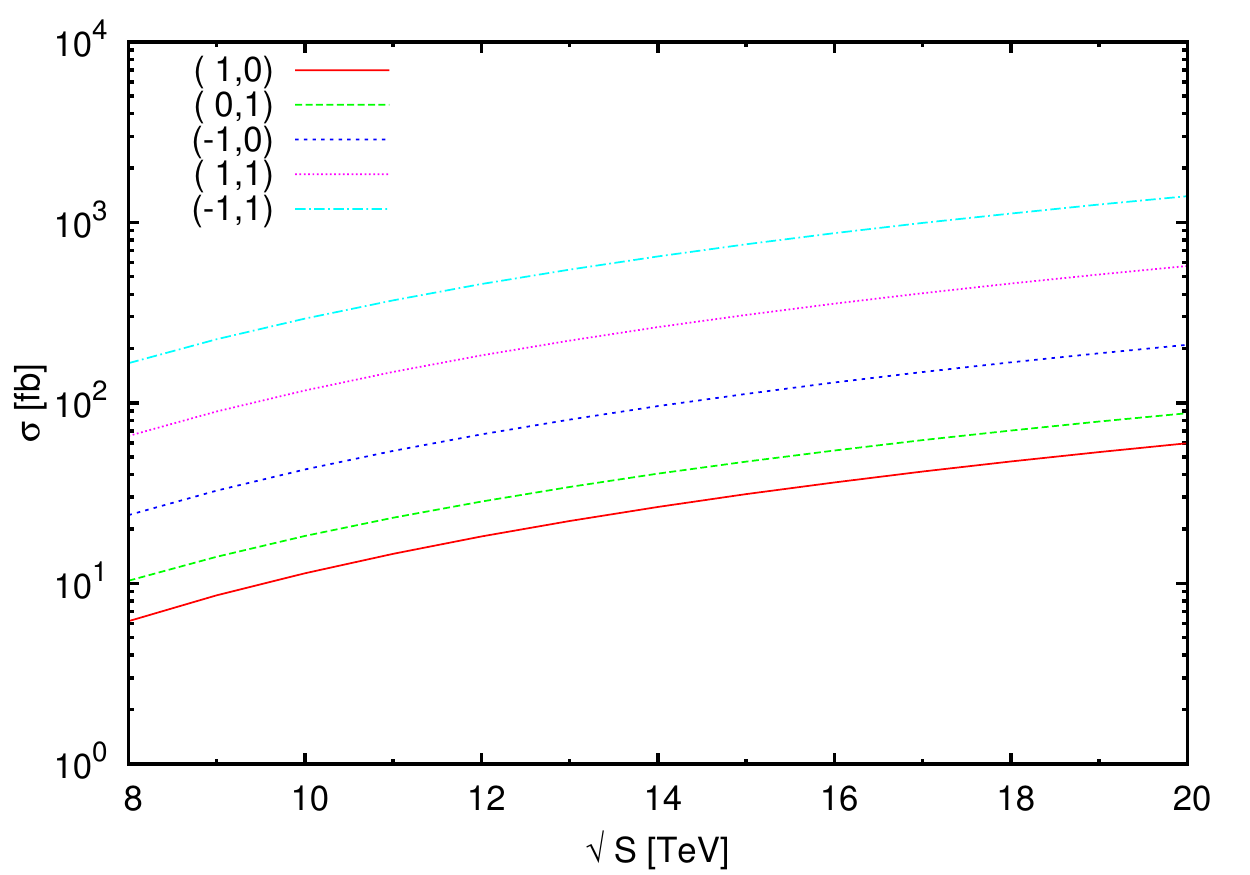}
  \captionsetup{width=.8\textwidth}
  \caption{Dependence of cross section on the collider center-of-mass energy for 
  various combinations of anomalous $ttH$ coupling parameters $(a,b)$.}
  \label{fig:cmeplot}
\end{minipage}
\end{figure}


We have adopted a semi-numerical approach to calculate the one-loop amplitude.
The quark loop traces for the box and triangle diagrams involving anomalous $ttH$ coupling 
are calculated using FORM in four dimensions~\cite{Vermaseren:2000nd}.
The one-loop tensor integrals which appear in the amplitude are reduced into one-loop scalars 
following the Oldenborgh and Vermaseren (OV) method~\cite{vanOldenborgh:1989wn}. 
The scalar integrals are calculated using the OneLOop package~\cite{vanHameren:2010cp}. 
We calculate helicity amplitudes numerically before squaring them to obtain the total 
and differential cross sections.
The numerical results presented in this section use CTEQ6L1 parton distribution functions~\cite{Nadolsky:2008zw}. 
We have taken $\mu = M_H$ ($=$ 125 GeV) as the common scale of renormalization and factorization. We have
not applied any kinematic cuts on the final state particles. 

\begin{table}[t]
 \begin{center}
  \begin{tabular}{|c|c|c|c|c|c|}
   \hline
   $\sqrt{\rm S}$& $\sigma_{(1,0)}$& $\sigma_{(0,\pm1)}$& $\sigma_{(-1,0)}$&  $\sigma_{(1,\pm1)}$
   &  $\sigma_{(-1,\pm1)}$ \\
   (TeV) & (fb)& (fb)& (fb)& (fb)& (fb)\\
   \hline
   8 & 6.18 & 10.34 & 23.89 & 65.58 & 165.89  \\
   \hline
   14 & 26.50 & 40.53 & 95.91 & 262.82 & 648.05  \\
   \hline
   33 & 167.51 & 234.94 & 567.27 & 1549.86 & 3719.29  \\
   \hline
  \end{tabular} 
 \end{center}
 \caption{$gg \to HH$ leading order hadronic cross sections for various combinations of parameters $(a,b)$.}
\end{table}

In Fig.~\ref{fig:scanplothh}, we can clearly  
see enhancement in the hadronic cross section due to the anomalous coupling parameters 
$a$ and $b$. {In pure pseudoscalar case ($a=0, b\ne0$), only the box diagrams contribute to the unpolarized cross section.}
For $a=-1$, the two diagrams in Fig~\ref{fig:gghh} interfere constructively 
leading to more than three fold increment in the cross section.
The cross section is indeed insensitive to any sign change in $b$.
We have further shown the cross sections for some benchmark values of $(a,b)$ as 
function of collider center-of-mass energy in Fig~\ref{fig:cmeplot}. For convenience, 
some of the numbers of interest are also given in the table 1. Although, these benchmark values
may not be realistic in the light of present LHC data on the Higgs-like particle, we consider them here 
for book keeping purpose. Apart from enhancing the production
cross section, these anomalous couplings also lead to characteristic changes in certain 
kinematic distributions. The distributions are presented for 14 TeV LHC.
\begin{figure}[t]
\centering
\begin{minipage}{.5\textwidth}
  \centering
  \includegraphics[width=7.5cm]{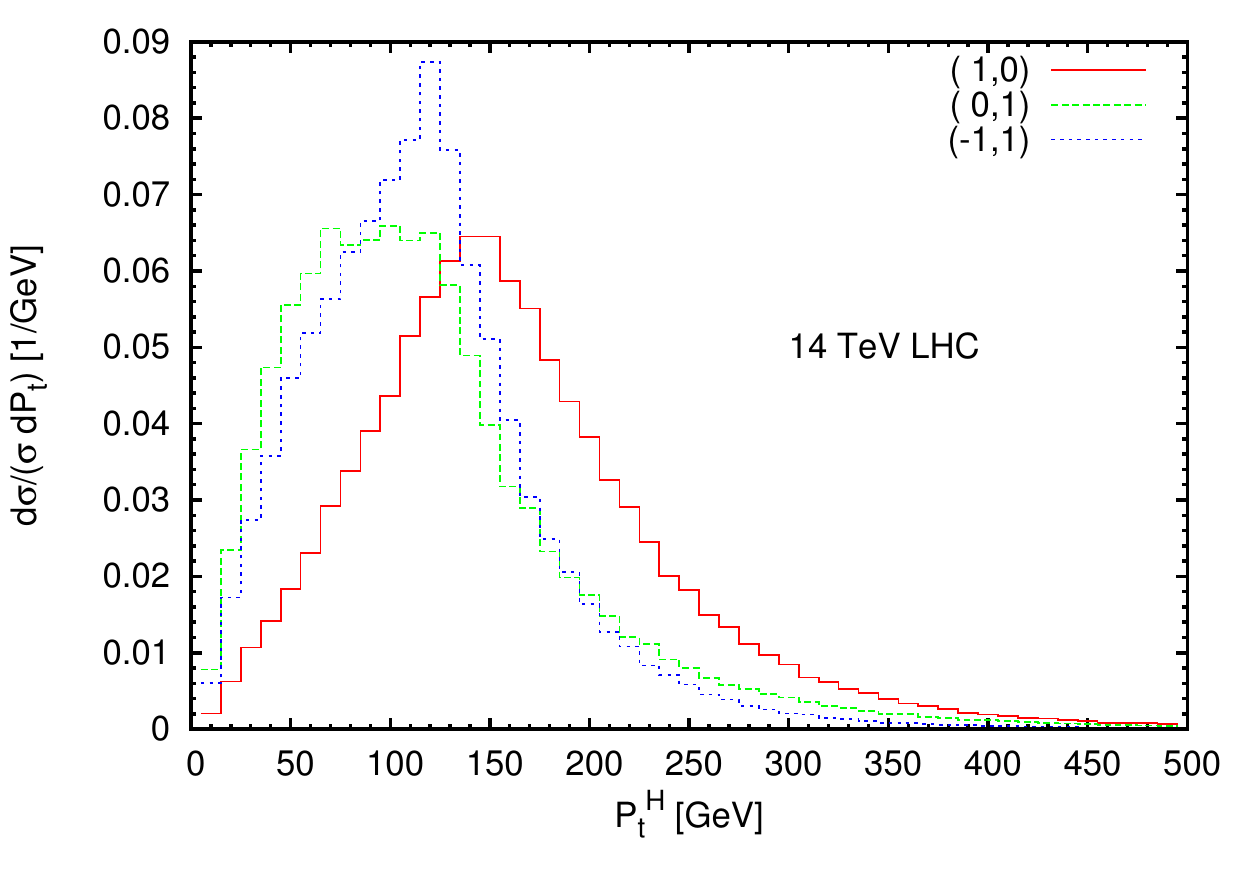}
  \captionsetup{width=.8\textwidth}
   \caption{Normalized $P_t$-distributions of the Higgs for various combinations 
          of anomalous $ttH$ coupling parameters $(a,b)$.}
  \label{fig:ptplot}
\end{minipage}%
\begin{minipage}{.5\textwidth}
  \centering
  \includegraphics[width=7.5cm]{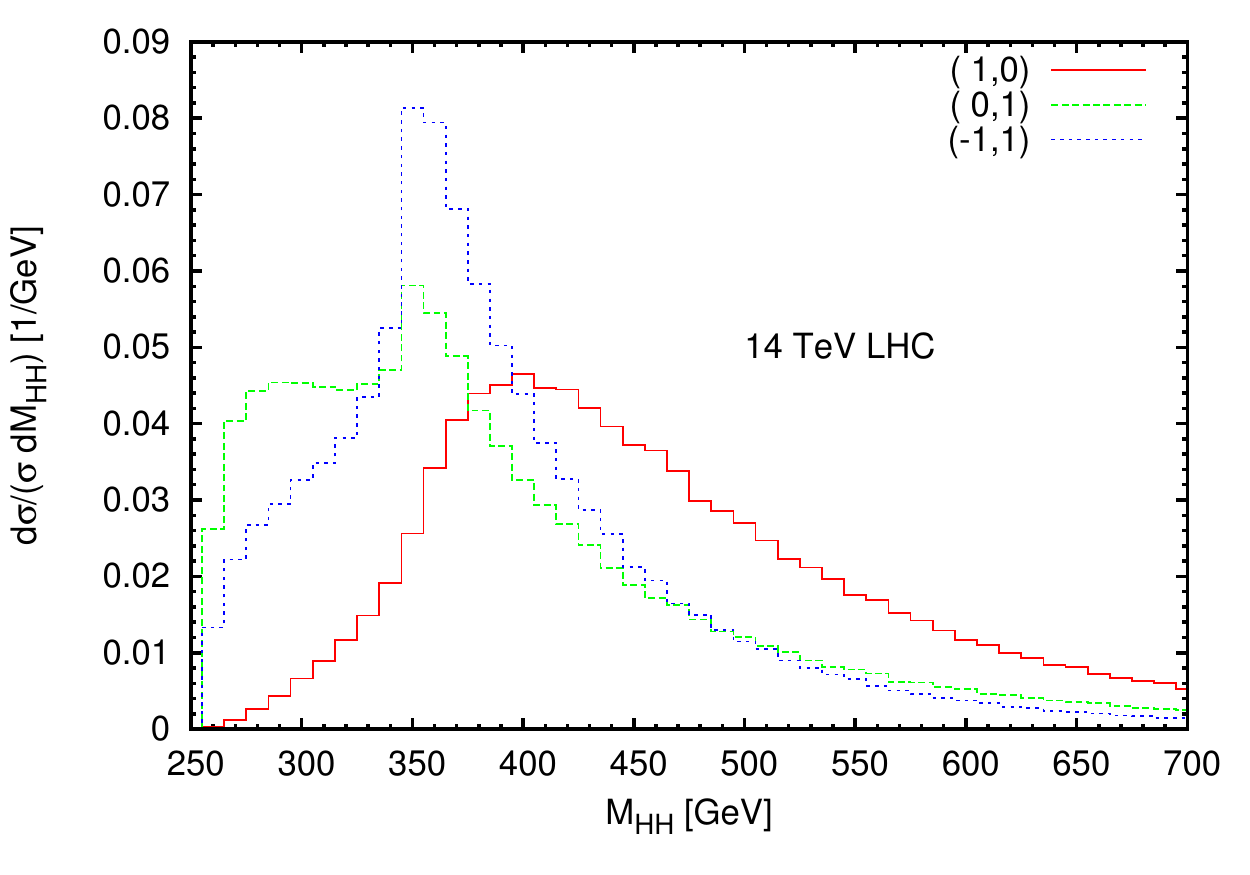}
  \captionsetup{width=.8\textwidth}
  \caption{Normalized two Higgs invariant mass distributions for various combinations 
          of anomalous $ttH$ coupling parameters $(a,b)$.}
  \label{fig:mhhplot}
\end{minipage}
\end{figure}
\begin{figure}[t]
\centering
\begin{minipage}{.5\textwidth}
  \centering
  \includegraphics[width=7.5cm]{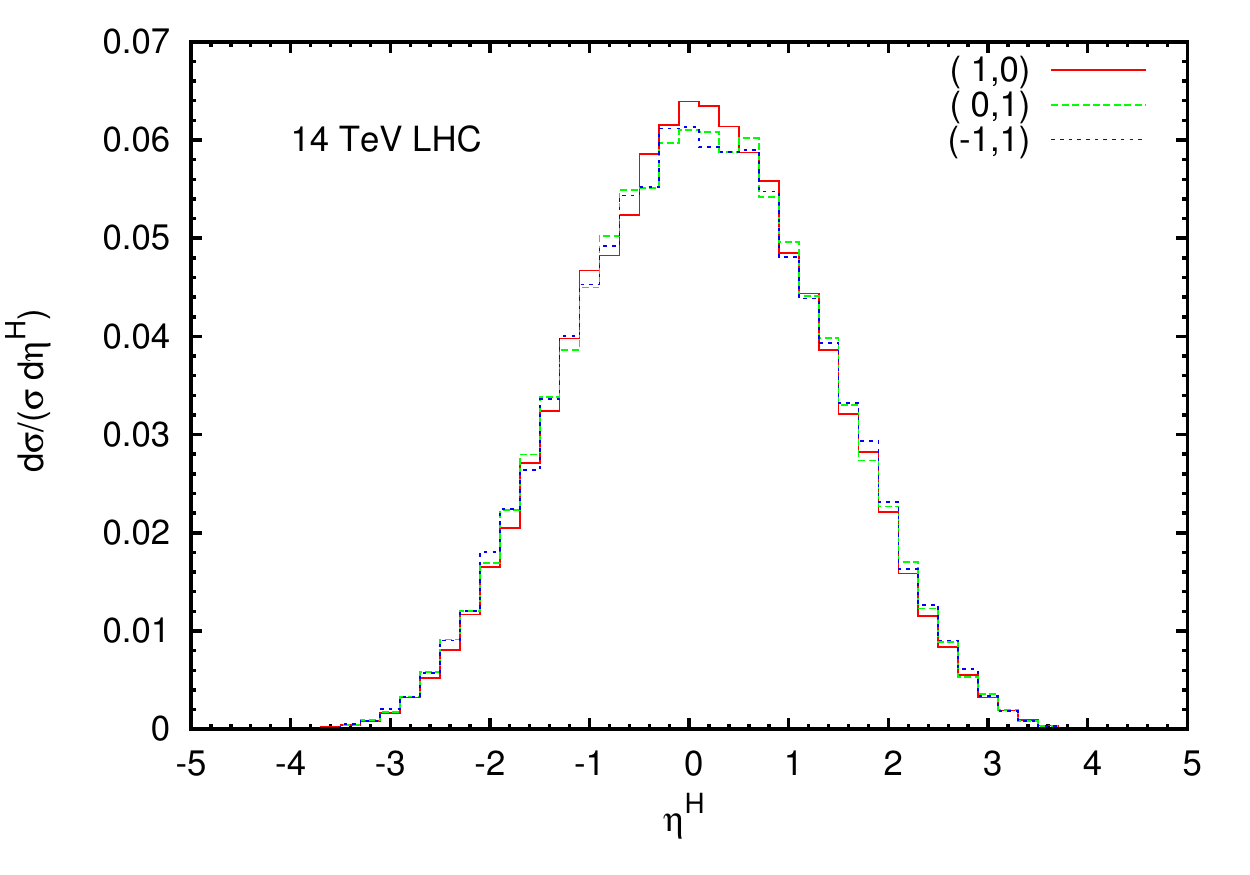}
  \captionsetup{width=.8\textwidth}
   \caption{Normalized rapidity distributions of the Higgs for various combinations 
            of anomalous $ttH$ coupling parameters $(a,b)$.}
  \label{fig:etaplot}
\end{minipage}%
\begin{minipage}{.5\textwidth}
  \centering
  \includegraphics[width=7.5cm]{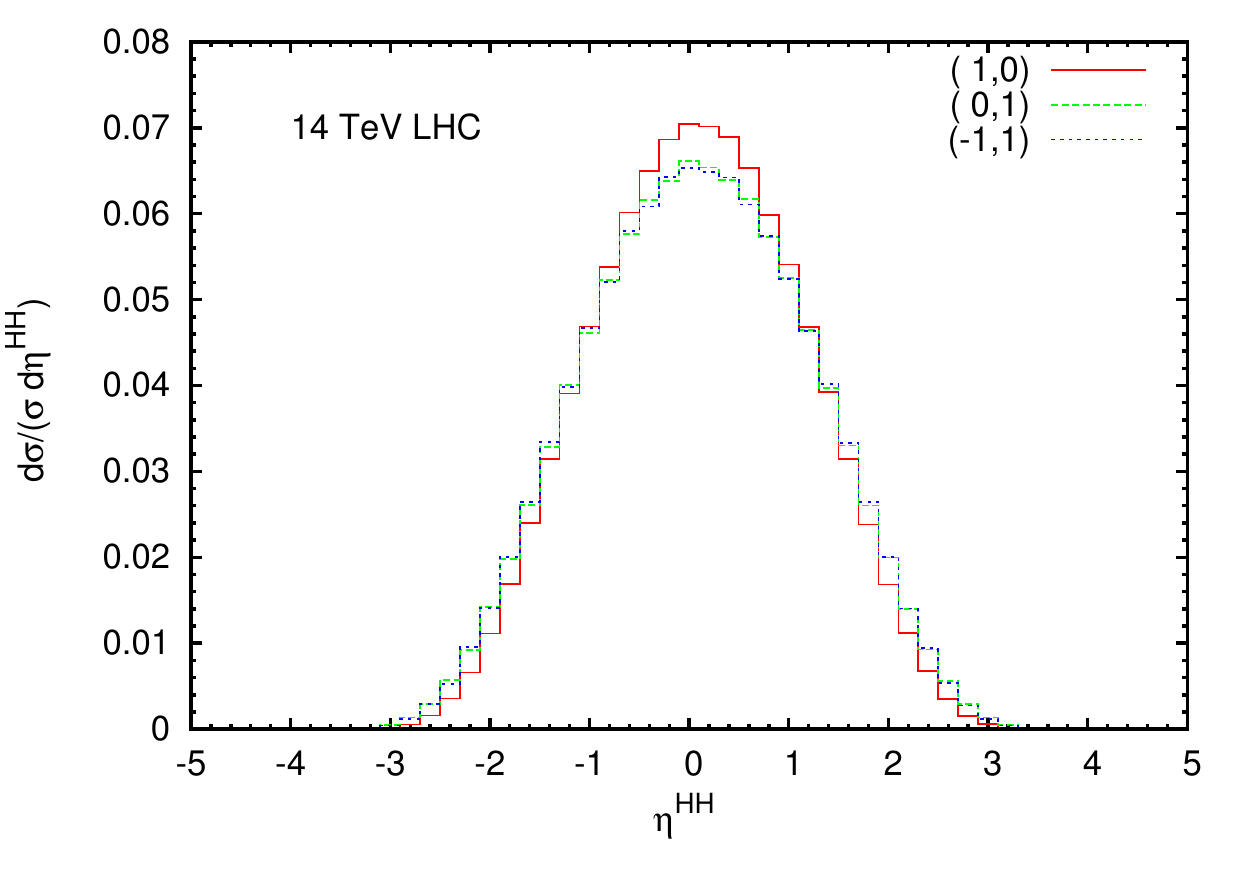}
  \captionsetup{width=.8\textwidth}
  \caption{Normalized rapidity distributions of the two Higgs system for various combinations 
          of anomalous $ttH$ coupling parameters $(a,b)$.}
  \label{fig:etahhplot}
\end{minipage}
\end{figure}

In Fig.~\ref{fig:ptplot}, we have compared the normalized transverse momentum distributions of Higgs
plotted for certain benchmark values of parameters $(a,b)$. We find that in presence of anomalous
couplings, the contribution from phase space region with $P_t^H$ below 150 GeV increases 
significantly. Similar conclusions are drawn from the invariant mass distributions ($M_{HH}$)
of the two Higgs bosons displayed in Fig~\ref{fig:mhhplot}. { The distributions start
at $M_{HH} = 2 M_H$ which is the production threshold for the two Higgs bosons in the final state. 
In the standard model case, there is an exact cancellation between the box and the triangle contributions in the large 
$m_t$ limit~\cite{Glover:1987nx}. This is clearly reflected in the low invariant mass region of the 
standard model distribution where large $m_t$ limit is a good approximation. Any deviation in the parameters
$(a,b)$ beyond standard model values dilutes this fine cancellation.}
The enhancement near $M_{HH} = 2m_t$ ($m_t =$ 172 GeV) threshold is also visible in these distributions. The rapidity 
distributions do not deviate much from the standard model case, see Figs.~\ref{fig:etaplot} and \ref{fig:etahhplot}. 
Similarly, the distribution corresponding to $\theta^\star_{HH}$ variable, 
discussed in Sec. 4 of Ref.~\cite{Baglio:2012np}, does not show any significant deviation.

 \begin{figure}[t]
 \includegraphics[width=3.15in]{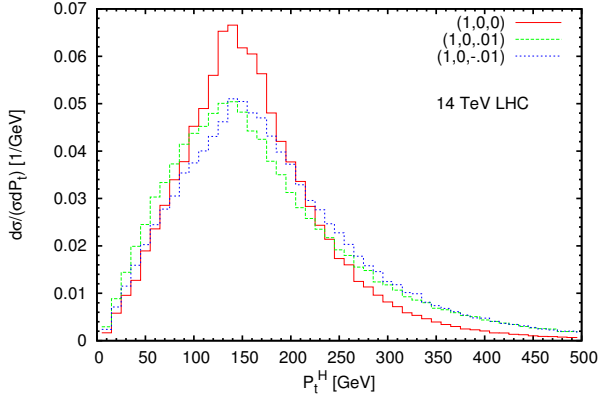}
 \includegraphics[width=3.15in]{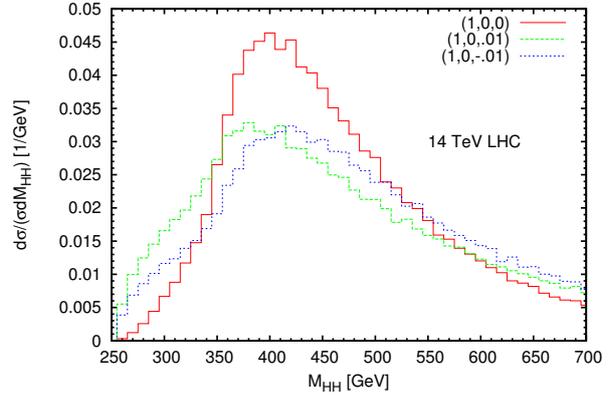}
 \caption{Effect of $ttHH$ coupling on normalized $P_t$ and invariant mass distributions in double Higgs production. At 14 TeV, for 
                 $c=-0.01$ and 0.01 the cross sections are 319.00 fb and 221.03 fb respectively. }
 \label{fig:Ctthh}
 \end{figure}

We would like to mention that in the context of the double Higgs production, the top-Higgs anomalous coupling can have more general 
features in addition to what is considered in Eq.~\ref{eq:tth_anml}. For example, in the effective Lagrangian approach, 
the operators shown in Eq.~\ref{eq:6DOP} also generate $ttHH$ contact interaction and it is related to the parameters of the $ttH$ coupling. Such contact 
interaction terms are also common in composite Higgs models~\cite{Contino:2012xk}. This new interaction can lead to drastic increment in the cross section 
of the double Higgs production, especially when the value of $\Lambda$ is quite low~\cite{Pierce:2006dh}. If we parametrize the $ttHH$ coupling 
factor by $(-m_t/v^2)c$ with a dimensionless parameter $c$, we find that for $c \gtrsim 0.001$ the effects are visible in both the cross section as well as in  
the transverse momentum and invariant mass distributions. As an illustration the normalized 
distributions for $c=\pm0.01$ are given in Fig.~\ref{fig:Ctthh} keeping $a$ and $b$ fixed at their standard model values.
Moreover, the anomalous couplings of the top quark with gluons and those of the Higgs boson with the gluons can 
also modify the Higgs pair production cross section at the LHC~\cite{Degrande:2012gr}. In presence of large number 
of free parameters, we loose the predictability and it becomes difficult to disentangle the effect of a specific parameter. 
To avoid this ambiguity we have not included any other anomalous coupling in our study.


\section{Constraints from LHC experiments
\label{Section:5}}
\indent
The LHC data on Higgs boson can be, in principle, used 
to constrain all those couplings which can affect the main production and/or decay 
channels of a single Higgs boson. However, we are interested in the couplings
of the Higgs with fermions and gauge bosons which might be sensitive to new physics. 
In this regard, $ttH$, $WWH$ and $ZZH$ couplings are the most relevant ones.
Just like the sources of anomalous term in case of $ttH$, 
similar higher dimensional operators could  modify $WWH$/$ZZH$ couplings as well. 
However, we note that such anomalous couplings of Higgs with gauge bosons are already constrained 
by the electroweak precision data.\footnote{
We note that there is no additional contribution to the Peskin--Takeuchi $S$, $T$, $U$ 
parameters due to the anomalous $ttH$ coupling at 1-loop level.}
Also, $ H \rightarrow WW^{*}$ and $ H \rightarrow ZZ^{*}$ are the two crucial
channels in which Higgs boson has been observed at the LHC. Therefore, these couplings get directly
constrained by the observed data. Hence, we do not intend to introduce any 
modifications to these couplings. In this section, we discuss the constraints on the anomalous top Yukawa
parameters from the latest results of Higgs searches at the LHC. 

The LHC experiments have collected data in the 
production channels which include the 
gluon fusion, the vector boson fusion, the Higgs-strahlung (associated production with 
a $W/Z$-boson), and the associated production with a pair of top quarks.
Under the existence of the top-Higgs anomalous coupling as shown in Eq.~(\ref{eq:tth_anml}), both the 
single Higgs production via gluon fusion and the Higgs production in association with $t\overline{t}$ are altered.
In addition to that, the partial decay widths of the Higgs to diphoton ($\Gamma_{H \to \gamma\gamma}$) 
and digluon ($\Gamma_{H \to gg}$) are deviated from those of the standard model values 
{($\Gamma^{\mathrm{SM}}_{H \to \gamma\gamma},\ \Gamma^{\mathrm{SM}}_{H \to gg}$)}.
The top-Higgs anomalous coupling also modifies the  $H \to Z\gamma$ decay width.
But this channel is hard to reconstruct and the constraints are still loose~\cite{ATLAS-CONF-2013-009,CMS-HIG-13-006}.
The branching ratio of this decay mode in standard model itself is small. Hence we do not expect
sizable deviation of the total Higgs decay
width coming from this channel. Therefore, we totally ignore the effects on $H \to Z\gamma$ due to the 
top-Higgs anomalous coupling {in this paper}.

In presence of anomalous top Yukawa coupling, the analytical expressions of the decay widths,
$\Gamma_{H \to gg}$ and $\Gamma_{H \to \gamma\gamma}$, are given by
\begin{align}
\Gamma_{H \to gg} &=
	\frac{G_F \alpha_s^2 M_H^3}{36\sqrt{2}\pi^3} \left\{ {\left| \frac{3}{4} a A_{1/2}(\tau_t) + \frac{3}{4} A_{1/2}(\tau_b) \right|^2}
	+ \left| \frac{3}{4} {\times 2b \frac{f(\tau_t)}{\tau_t}} \right|^2 \right\}, \label{Htogg_deviation} \\
\Gamma_{H \to \gamma\gamma} &=
	\frac{G_F \alpha^2 M_H^3}{128\sqrt{2}\pi^3}
	\left\{ \left| A_1(\tau_W) + a N_C Q_t^2 A_{1/2}(\tau_t) {+
	N_C Q_b^2 A_{1/2}(\tau_b)} \right|^2 +
	\left| N_C Q_t^2 {\times 2b \frac{f(\tau_t)}{\tau_t}} \right|^2 \right\}, \label{Htogammagamma_deviation}
\end{align}
with the functions of $\tau_i$ as in Ref.~\cite{Djouadi:2005gj}, which is defined as $\tau_i \equiv M_H^2/4M_i^2$,
\begin{align}
f(\tau_i) &= 
	\begin{cases} \displaystyle \arcsin^2({\sqrt{\tau_i}}) & \tau_i \leq 1, \\
	\displaystyle {-\frac{1}{4} \left[ \log{\frac{1+\sqrt{1-\tau_i^{-1}}}{1-\sqrt{1-\tau_i^{-1}}}} -i\pi \right]^2} & \tau_i > 1, \end{cases} \\
A_{1/2}(\tau_i) &= \frac{2}{\tau_i^2} \left[ \tau_i + (\tau_i-1) f(\tau_i) \right], \\
A_{1}(\tau_i) &= - \frac{1}{\tau_i^2} \left[ 2\tau_i^2 + 3\tau_i + 3(2\tau_i -1)f(\tau_i) \right].
\end{align}
Here $G_F$ is the Fermi constant, $\alpha_s$ and $\alpha$ are the fine structure constants for QCD and QED, 
and $N_C$, $Q_t$($Q_b$) represent the QCD color factor and electric charge of the top(bottom) quark, respectively.
Note that we also include the contribution from the bottom quark since the corresponding loop 
function $A_{1/2}(\tau_b)$ is non-negligible.

\begin{figure}[t]
\begin{center}
   \includegraphics[width=0.32\columnwidth]{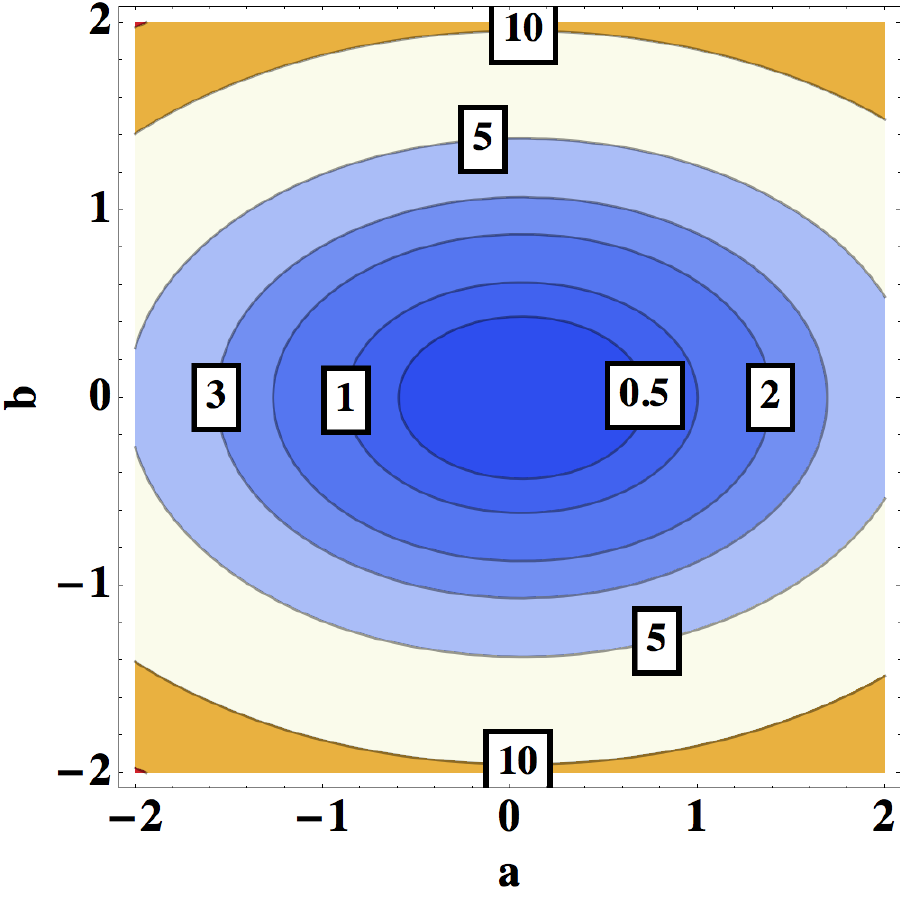}
   \includegraphics[width=0.32\columnwidth]{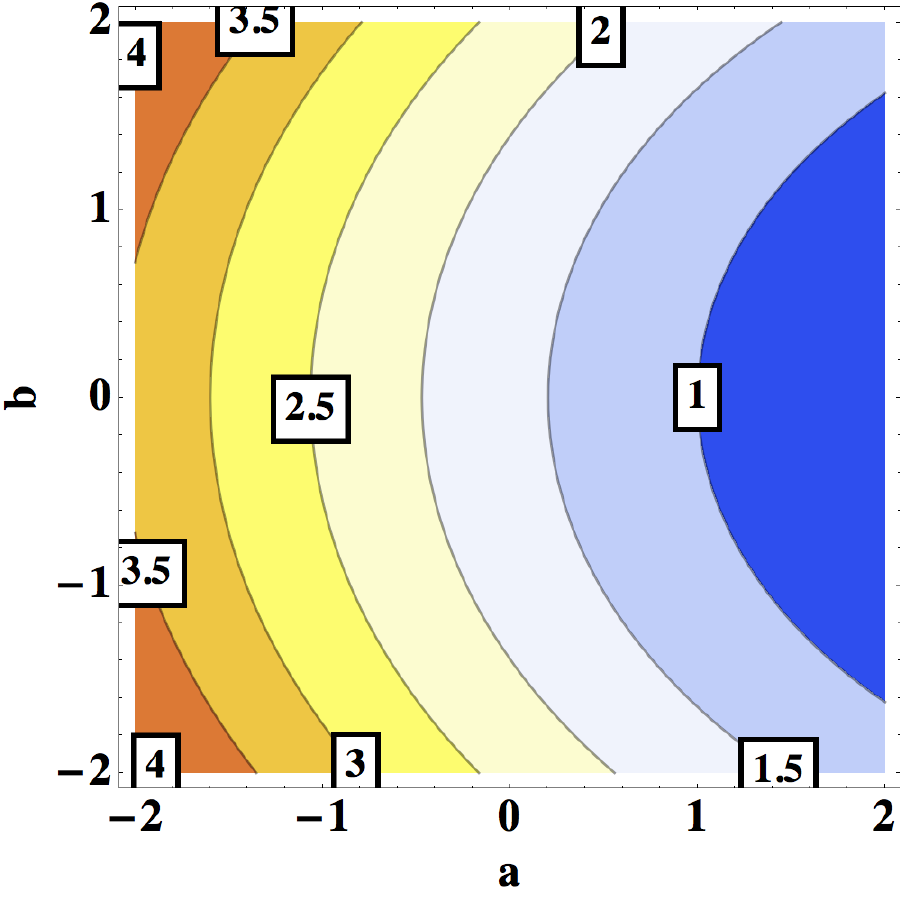}
   \includegraphics[width=0.32\columnwidth]{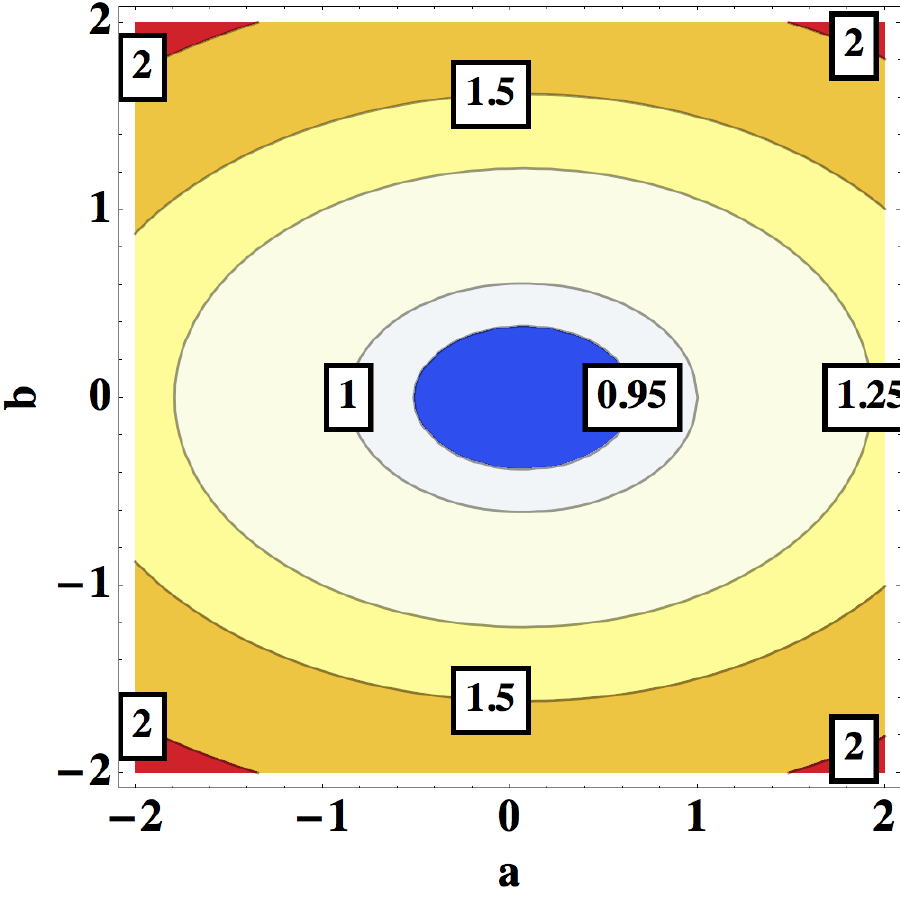}
\caption{The deviations of $\Gamma_{H \to gg}/\Gamma_{H \to gg}^{\mathrm{SM}}$(left), 
$\Gamma_{H \to \gamma\gamma}/\Gamma_{H \to \gamma\gamma}^{\mathrm{SM}}$(center) and 
$\Gamma_H/\Gamma_H^{\mathrm{SM}}$(right) as functions of $a$ and $b$, respectively.
}
\label{fig:widths_deviation}
\end{center}
\end{figure}

As mentioned earlier,  $\Gamma_{H \to gg}$ and $\Gamma_{H \to \gamma\gamma}$ and hence, 
total Higgs decay width $\Gamma_H$ change from their standard model values due to the presence 
of modified $ttH$ coupling.
The following ratio is suitable for evaluating this effect:
\begin{align}
\frac{\Gamma_H}{\Gamma_H^{\mathrm{SM}}} &=
	\mathrm{Br}_{H \to \mathrm{others}}^{\mathrm{SM}} +
	\frac{\Gamma_{H \to gg}}{\Gamma_{H \to gg}^{\mathrm{SM}}} \mathrm{Br}_{H \to gg}^{\mathrm{SM}} +
	\frac{\Gamma_{H \to \gamma\gamma}}{\Gamma_{H \to \gamma\gamma}^{\mathrm{SM}}} \mathrm{Br}_{H \to \gamma\gamma}^{\mathrm{SM}}, \label{totalgamma_deviation}
\end{align}
where $\mathrm{Br}_{H \to \mathrm{others}}^{\mathrm{SM}} = 0.913$, $\mathrm{Br}_{H \to gg}^{\mathrm{SM}} = 0.085$
and $\mathrm{Br}_{H \to \gamma\gamma}^{\mathrm{SM}} = 0.002$ are the branching ratios at around $M_H = 125\,\mathrm{GeV}$
in the standard model~\cite{Giardino:2012ww}.
We assume that the $K$-factors are the same as those in the standard model and are dropped  in Eq.~(\ref{totalgamma_deviation}).
Figure~\ref{fig:widths_deviation} shows the deviations of $\Gamma_{H \to gg}/\Gamma_{H \to gg}^{\mathrm{SM}}$,
$\Gamma_{H \to \gamma\gamma}/\Gamma_{H \to \gamma\gamma}^{\mathrm{SM}}$ and $\Gamma_H/\Gamma_H^{\mathrm{SM}}$ as
functions of the top-Higgs anomalous parameters $a$ and $b$.
The three ratios are more sensitive to the parameter $b$ compared to $a$
because of the largeness of 
the loop function, $A_{1/2}(\tau_t) \simeq 1.4$ and $2 f(\tau_t)/\tau_t \simeq 2.1$.
In negative region of $a$, due to the constructive interference of  $W$ and the quark loop contributions,
the deviation in $\Gamma_{H \to \gamma\gamma}/\Gamma_{H \to \gamma\gamma}^{\mathrm{SM}}$
turns out to be significant.
Because the value of $\mathrm{Br}_{H \to gg}^{\mathrm{SM}} = 0.085$ is not so small, the ratio 
of the total width $\Gamma_H/\Gamma_H^{\mathrm{SM}}$ {receives} a sizable modification in the 
region where $\Gamma_{H \to gg}/\Gamma_{H \to gg}^{\mathrm{SM}}$ is large.

Now, we address the deviations in cross sections of the single Higgs production processes due to the top-Higgs
anomalous coupling. The leading order cross section in gluon fusion channel can be evaluated from:
\begin{equation}
\hat{\sigma}_{gg \to H} = \frac{\pi^2}{8M_H} \Gamma_{H \to gg} \delta(\hat{s} - M_H^2),
\label{sigma_pptoH_partonlevel}
\end{equation}
where the hat symbol indicates that {it is a parton level value}.
The form in Eq.~(\ref{sigma_pptoH_partonlevel}) {suggests that, at the hadron level,} the parton-distribution part 
should be factorized and we can conclude the following relation:
\begin{equation}
\frac{{\sigma}_{gg \to H}}{{\sigma}_{gg \to H}^{\mathrm{SM}}} =
	\frac{\Gamma_{H \to gg}}{\Gamma_{H \to gg}^{\mathrm{SM}}}.
	\label{cross_section_deviation}
\end{equation}
Therefore, the left most plot in Fig.~\ref{fig:widths_deviation} also represents deviations in $gg \to  H$ cross section in 
presence of anomalous coupling parameters $a$ and $b$.
For calculating the deviation $\sigma_{pp \to t\overline{t}H}/\sigma_{pp \to t\overline{t}H}^{\mathrm{SM}}$, we implement the 
anomalous coupling with the help of FeynRules~\cite{Christensen:2008py} and generate a Universal FeynRules Output (UFO) model 
file~\cite{Degrande:2011ua} for Madgraph 5~\cite{Alwall:2011uj}.
{The left contour plot of Fig.~\ref{fig:pptottH} shows the ratio $\sigma_{pp \to t\overline{t}H}/\sigma_{pp \to t\overline{t}H}^{\mathrm{SM}}$ 
as a function of $a$ and $b$ at $\sqrt{s} = 8\,\mathrm{TeV}$, which is symmetric under $a \to -a$ or $b \to -b$ and the effect of $b$ is 
subleading in contradiction to $gg \to H$, $H \to gg$ and $H \to \gamma\gamma$.}
We use the CTEQ6L1 parton distribution function for calculating the cross section.
Both the renormalization and factorization scales have been set at ($2 m_t + M_H$).

{The ATLAS and the CMS experiments have published the inclusive results of $H \to \gamma\gamma$, $H \to ZZ^{*} \to 4\ell$ and 
$H \to WW^{*} \to 2\ell2\nu$ for each category tagging their decays~\cite{ATLAS-CONF-2013-012,ATLAS-CONF-2013-013,ATLAS-CONF-2013-030,CMS_PAS_HIG-13-001,CMS_PAS_HIG-13-002,CMS_PAS_HIG-13-003}, where all the production channels are considered.}
Also,  $H \to b\overline{b}$ after the production through the vector boson fusion~\cite{CMS_PAS_HIG-13-011} 
and the Higgs-strahlung~\cite{ATLAS-CONF-2013-079,CMS_PAS_HIG-13-012} have been reported.
We can put a bound on the $(a,b)$-plane after executing a global analysis based on the above data.\footnote{
Lots of works have been done before and after the Higgs discovery.
See e.g., Refs in~\cite{recent_globalanalysis} for recent status.}
On the other hand, the signal strength of $pp \to t\overline{t}H$ (subsequently, $H \to \gamma\gamma$ or $H \to b\overline{b}$) is 
now constrained {at the LHC}. The ATLAS have claimed that at the $95\%$ CL the observed upper limits from 
$H \to \gamma\gamma$ and $H \to b\overline{b}$ are $5.3$~\cite{ATLAS-CONF-2013-080} and $13.1$~\cite{ATLAS-CONF-2012-135} respectively, 
while the CMS counterparts are $5.4$ $(H \to \gamma\gamma)$~\cite{CMS_PAS_HIG-13-015} and $5.8$ ($H \to b\overline{b}$)~\cite{CMS_HIG-12-035}.
Since, the top-Higgs anomalous coupling can modify these sequences of production and decay, additional restrictions on $a$ and $b$ can be imposed.
Due to the large uncertainties, we do not use these data in our global analysis and 
separately examine a bound from this channel without considering errors seriously. The right plot in Fig~\ref{fig:pptottH} 
represents the regions where the results are consistent with the CMS observations; $\mu_{pp \to t\overline{t}H, H \to \gamma\gamma} \leq 5.4$ ({cyan}) or $\mu_{pp \to t\overline{t}H, H \to b\overline{b}} \leq 5.8$ ({magenta}).
{The tendency of the two constraints can be understood from the properties of the three fractions $\Gamma_{H \to gg}/\Gamma_{H \to gg}^{\mathrm{SM}}$, $\Gamma_{H \to \gamma\gamma}/\Gamma_{H \to \gamma\gamma}^{\mathrm{SM}}$ and $\Gamma_H/\Gamma_H^{\mathrm{SM}}$ which we discussed before.}
The purple area is the superposition of the two allowed regions.

In order to take into account the difference in the production processes in our global analysis, 
we employ the following weight used in Refs.~\cite{Abe:2012eu,Kakuda:2013kba}:
\begin{equation}
\epsilon^{I,X}_{f} = \frac{a^{I,X}_{f} \sigma^{\mathrm{SM}}_{X}}{\sum_{Y} a^{I,Y}_{f} \sigma^{\mathrm{SM}}_{Y}},
\end{equation}
where $X$ and $I$ are indices to distinguish the production channels and event categories in the 
decay $H \to f$, $\sigma_X^{\mathrm{SM}}$ is the single Higgs production cross section of the 
channel $X$ in the standard model, and $a^{I,Y}_{f}$ means acceptances.
After ignoring the deviations in acceptances originating from effects of new physics, we can 
identify the weight factor $\epsilon^{I,X}_{f}$ as the fractions of expected signal events 
from the five production processes, whose details are {provided} in 
Refs.~\cite{ATLAS-CONF-2013-012,ATLAS-CONF-2013-013,ATLAS-CONF-2013-030,CMS_PAS_HIG-13-001,CMS_PAS_HIG-13-002,CMS_PAS_HIG-13-003,CMS_Moriond}
and summarized in section 3 of Ref.~\cite{Kakuda:2013kba}.
Note that the simple relation $\sum_X \epsilon^{I,X}_{f} = 1$ holds.
After the set $\{ \epsilon^{I,X}_{f} \}$ is ready in the decay $H \to f$, the signal strength can be written down as follows:
\begin{equation}
\mu^{I}_{H \to f} = \sum_{X} \epsilon^{I,X}_{f} \frac{\sigma_X}{\sigma_X^{\mathrm{SM}}}
\frac{\mathrm{Br}_{H \to f}}{\mathrm{Br}_{H \to f}^{\mathrm{SM}}}
= \sum_{X} \epsilon^{I,X}_{f} \frac{\sigma_X}{\sigma_X^{\mathrm{SM}}}
\frac{\Gamma_{H \to f}}{\Gamma_{H \to f}^{\mathrm{SM}}}
\frac{\Gamma_H^{\mathrm{SM}}}{\Gamma_H},
\end{equation}
where $\sigma_X$ represents the Higgs production cross section of the process $X$ with the top-Higgs 
anomalous coupling, and $\mathrm{Br}^{(\mathrm{SM})}_{H \to f} = \Gamma^{(\mathrm{SM})}_{H \to f}/\Gamma^{(\mathrm{SM})}_{H}$ 
is the branching ratio of the Higgs decay channel {$H \to f$} (in the standard model).
The possible deviations via loop corrections of {the ratios}, ${\Gamma_{H \to f}}/{\Gamma_{H \to f}^{\mathrm{SM}}}$ and 
${\Gamma_H^{\mathrm{SM}}}/{\Gamma_H}$ are already evaluated in Eqs.~(\ref{Htogg_deviation}), (\ref{Htogammagamma_deviation}) 
and (\ref{totalgamma_deviation}).
{We mention that all the other ratios have no deviation from the standard model.
As mentioned earlier, the ratio ${\sigma_X}/{\sigma_X^{\mathrm{SM}}}$ deviates from one only in the gluon 
fusion production channel and in $pp \to t\overline{t} H$ production channel.

\begin{figure}[t]
\begin{center}
   \includegraphics[width=0.32\columnwidth]{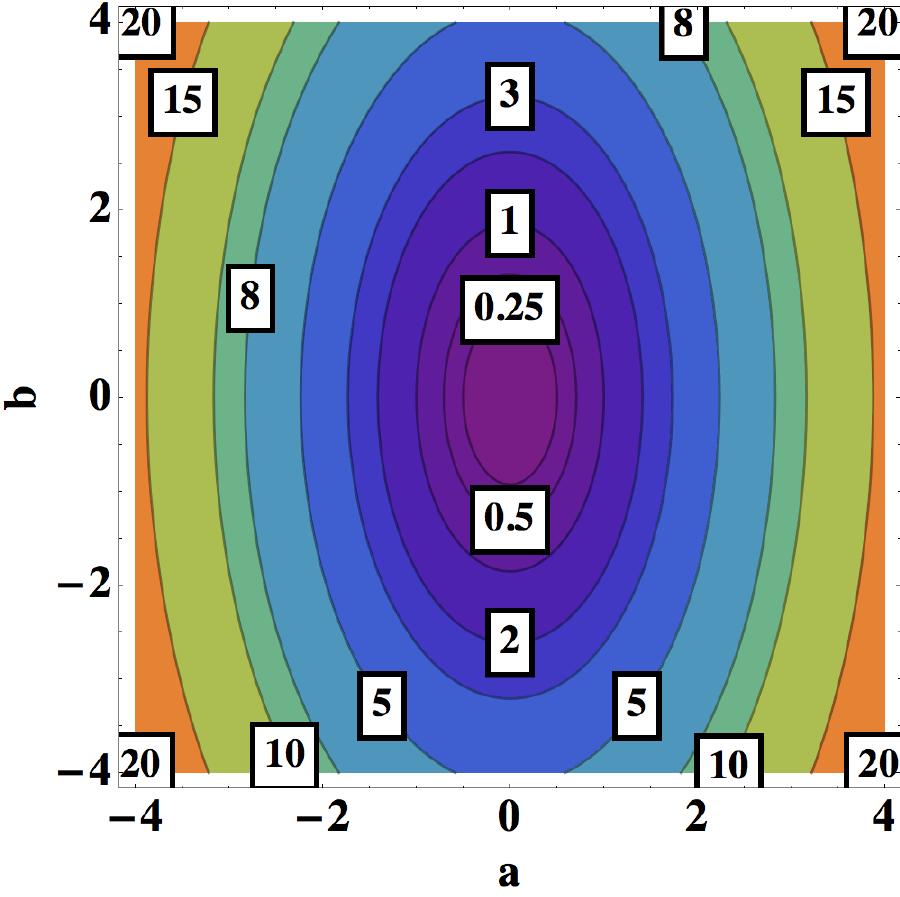} \hspace{10mm}
   \includegraphics[width=0.32\columnwidth]{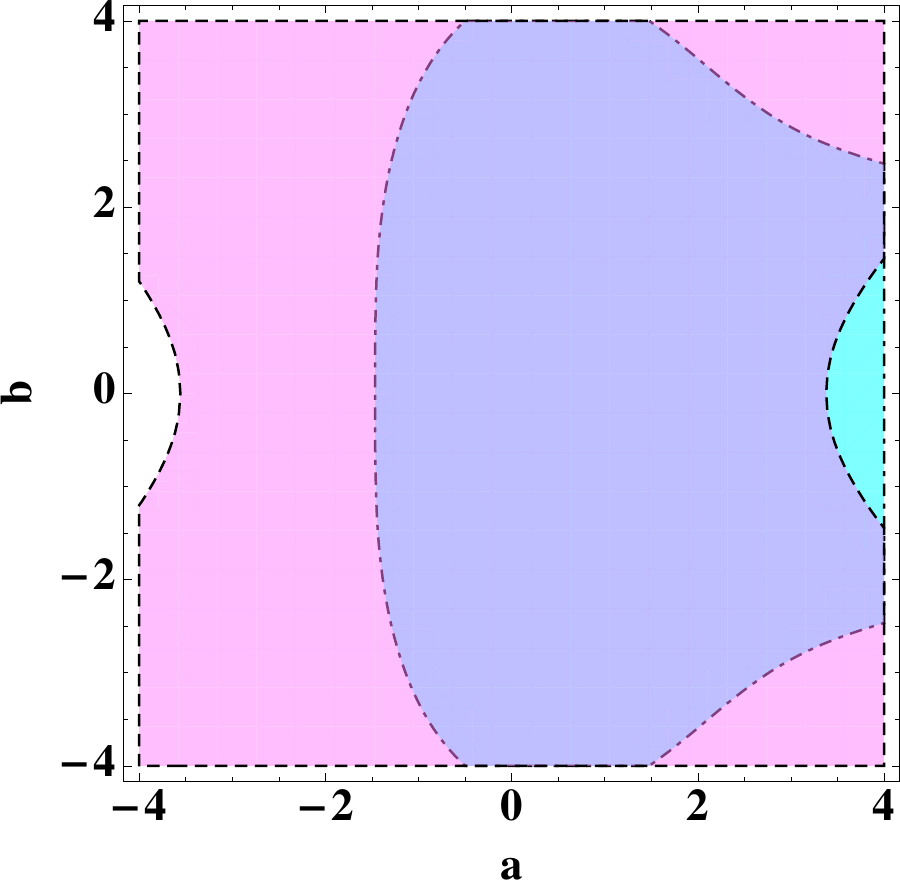}
\caption{
{\it Left}\,: $\sigma_{pp \to t\overline{t}H}/\sigma_{pp \to t\overline{t}H}^{\mathrm{SM}}$ as a function of $a$ and $b$ at $\sqrt{s} = 8\,\mathrm{TeV}$.
{\it Right}\,: parameter regions being consistent with the CMS observations; $\mu_{pp \to t\overline{t}H, H \to \gamma\gamma} \leq 5.4$ ({cyan})~\cite{CMS_PAS_HIG-13-015} or
$\mu_{pp \to t\overline{t}H, H \to b\overline{b}} \leq 5.8$ ({magenta})~\cite{CMS_HIG-12-035}. The purple area is the superposition of the two allowed regions.
}
\label{fig:pptottH}
\end{center}
\end{figure}



\begin{table}[t]
\begin{center}
\begin{tabular}{|c||c|c|}
\hline
Type & Signal strength & Reference \\ \hline \hline
ATLAS, VH($bb$), 0 lepton & $0.5^{+0.9}_{-0.9}$ &  \\
ATLAS, VH($bb$), 1 lepton & $0.1^{+1.0}_{-1.0}$ & \cite{ATLAS-CONF-2013-079} \\
ATLAS, VH($bb$), 2 lepton & $-0.4^{+1.5}_{-1.4}$ &  \\ \hline
CMS, $Z(\nu\nu)H(bb)$ & $1.04 \pm 0.77$ & \\
CMS, $Z(\ell^+ \ell^-)H(bb)$ & $0.82 \pm 0.97$ & \cite{CMS_PAS_HIG-13-012} \\
CMS, $Z(\ell\nu)H(bb)$ & $1.11 \pm 0.87$ & \\ \hline
CMS, H(bb) from VBF & $0.7 \pm 1.4$ & \cite{CMS_PAS_HIG-13-011}
\\ \hline
\end{tabular}
\caption{
Details of $7$ observables of $H \to b\overline{b}$ after the production 
through the vector boson fusion and the Higgs-strahlung.
}
\label{table:LHCobservables}
\end{center}
\end{table}

\begin{figure}[t]
\begin{center}
   \includegraphics[width=0.5\columnwidth]{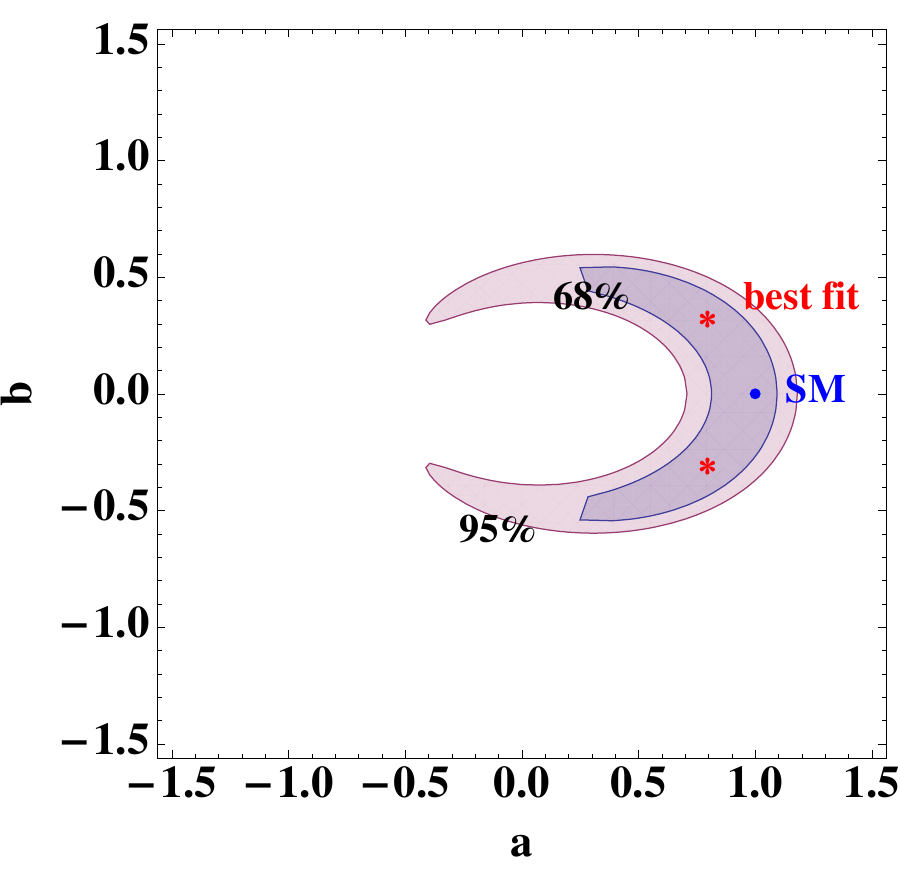}
\caption{
$68\%$ and $95\%$ CL allowed regions of the global analysis.
The red asterisks and the blue point represent the best-fit point and the standard model point, respectively.
}
\label{fig:globalanalysis}
\end{center}
\end{figure}

Next, we perform a $\chi^2$ analysis with the ATLAS and the CMS results with the $\chi^2$ function {defined as}
\begin{equation}
\chi^2 = 
	\sum_f \sum_I \left( \frac{\mu^I_{H \to f} - \hat{\mu}^I_f}{\hat{\sigma}^I_f} \right)^2.
\end{equation}
 We assume every experimental result follows Gaussian distribution ({$\hat{\mu}^I_f \pm \hat{\sigma}^I_f$}) and ignore 
 the correlations among the event categories, which are not yet published.
{When an error is asymmetric, we adopt its simple average as the value of the corresponding $\hat{\sigma}^I_f.$}
We use $42$ observables of $H \to \gamma\gamma$~\cite{ATLAS-CONF-2013-012,CMS_PAS_HIG-13-001}, 
$H \to ZZ^{*} \to 4\ell$~\cite{ATLAS-CONF-2013-013,CMS_PAS_HIG-13-002}, 
and $H \to WW^{*} \to 2\ell 2\nu$~\cite{ATLAS-CONF-2013-030,CMS_PAS_HIG-13-003}, whose details are summarized in section 3 of Ref.~\cite{Kakuda:2013kba}, and $7$ ones of $H \to b\overline{b}$ after the production through the vector boson fusion~\cite{CMS_PAS_HIG-13-011} and the Higgs-strahlung~\cite{ATLAS-CONF-2013-079,CMS_PAS_HIG-13-012}, where we can find the values in table~\ref{table:LHCobservables}.
We note that the number of the inputs is $49$ in total.
The $68\%$ and $95\%$ CL allowed regions are shown {in Fig.~\ref{fig:globalanalysis}}, where the best-fit point 
(global minimum of $\chi^2$) is found at {$(a,b) = (0.796, \pm 0.315)$ with $\chi^2_{\mathrm{min}} = 36.0$}.
The large area near the  point ($0,0$) in the ($a,b$)-plane is disfavored because the dominant single Higgs production via gluon 
fusion is suppressed much and this is {in contradiction} to the (inclusive) experimental results.
The anomalous coupling $b$ is more restricted than $a$ since deviation of $b$ plays the primary role in the single Higgs 
production and its decay. We mention that, after combining the result of our global analysis with the constraint 
from the $pp \to t\overline{t}H$, which is shown in Fig.~\ref{fig:pptottH}, the favored region does not change.

We should mention that this estimation is rather crude  because of lack of error consideration.
We hope that we can be more confident on our results after accumulation of further data in $pp \to t\overline{t}H$ process in the near future.
We should also emphasize that we only consider anomalous couplings in the top Yukawa sector in the global analysis.
After introducing deviations in other couplings, the result might get modified.

\section{Summary and Discussions
\label{Section:6}}

In this paper, we consider anomalous $ttH$ coupling and explore its effects on the Higgs 
pair production at the LHC. The term `anomalous' is an 
indication of possible new physics beyond the standard model. This anomalous coupling
describes  that the standard model top-Higgs Yukawa coupling is deviated by a scale factor `$a$'
along with an extra pseudo-scalar type coupling parameterized by `$b$'. For definiteness,  
we do not consider possible anomalous couplings of the Higgs with other fermions/bosons.

In section 4, we have considered $O(1)$ deviations in the anomalous coupling parameters $a$ and $b$ 
from their standard model values. With such deviations one finds large enhancement in the Higgs 
pair production cross section. But this deviation can also contrast already gained knowledge on Higgs
couplings on the basis of analyzed data at the LHC. Therefore, we constrain the parameter space
by doing a global analysis based on data released by the ATLAS and CMS and show the allowed region
in Fig.~\ref{fig:globalanalysis}. The best-fit values obtained for the anomalous parameters are $(0.8,\pm 0.3)$.
Both the Higgs production via gluon fusion and its decay to digluon are affected more by $b$ than by $a$. 
{On the other hand, in the Higgs decay to diphoton, the deviation in $a$ also plays an important role.}
 We find that non-zero values of the pseudoscalar coupling parameter $b$ are consistent with the data, but
$a=-1$ case is completely ruled out at $95\%$ CL. For $a=1$, the parameter $b$ is allowed to take any value between $-0.4$ to $+0.4$.
Tight constrains on anomalous parameters indicate the consistency of LHC data with the standard model predictions.
We would like to reiterate that the results of global analysis is not a sophisticated one. Once we introduce
anomalous couplings of Higgs with other fermions/bosons, the allowed region of parameter space is likely to change.
 \begin{table}[t]
 \begin{center}
  \begin{tabular}{|c|c|c|c|c|c||c|c|}
   \hline
   $\sqrt{\rm S}$& $\sigma_{(1,0,1)}$&$\sigma_{(1.2,0,1)}$& $\sigma_{(0,\pm0.6,1)}$& $\sigma_{(-0.4,\pm0.4,1)}$&  $\sigma_{(0.8,\pm0.3,1)}$&  
   $\sigma_{(1,0,0)}$&  $\sigma_{(1,0,2)}$ \\
   (TeV) & (fb)& (fb)& (fb)& (fb)& (fb)& (fb)& (fb) \\
   \hline
    8 &  6.18 & 14.70 &  2.67 &   7.19 &   4.84 &  13.18 &  2.87 \\
   \hline
   14 &  26.50 & 62.51 & 10.26 &  27.83 &  21.57 &  54.22 & 12.76 \\
   \hline
   33 &  167.51 & 391.56 & 57.91 & 157.88 & 122.02 & 328.67 & 83.85 \\
   \hline
  \end{tabular} 
 \end{center}
 \caption{Higgs pair production cross sections for benchmark values of $(a,b,\kappa)$ consistent with 
          the LHC data. The parameter $\kappa$ is the scale factor for the trilinear Higgs coupling 
          defined below in the text. Numbers in 1st column stand for the standard model value, while 
          the 5th column correspond to the best-fit value of the parameters.}
\end{table}
\begin{figure}[t]
\centering
\begin{minipage}{.5\textwidth}
  \centering
  \includegraphics[width=7.5cm]{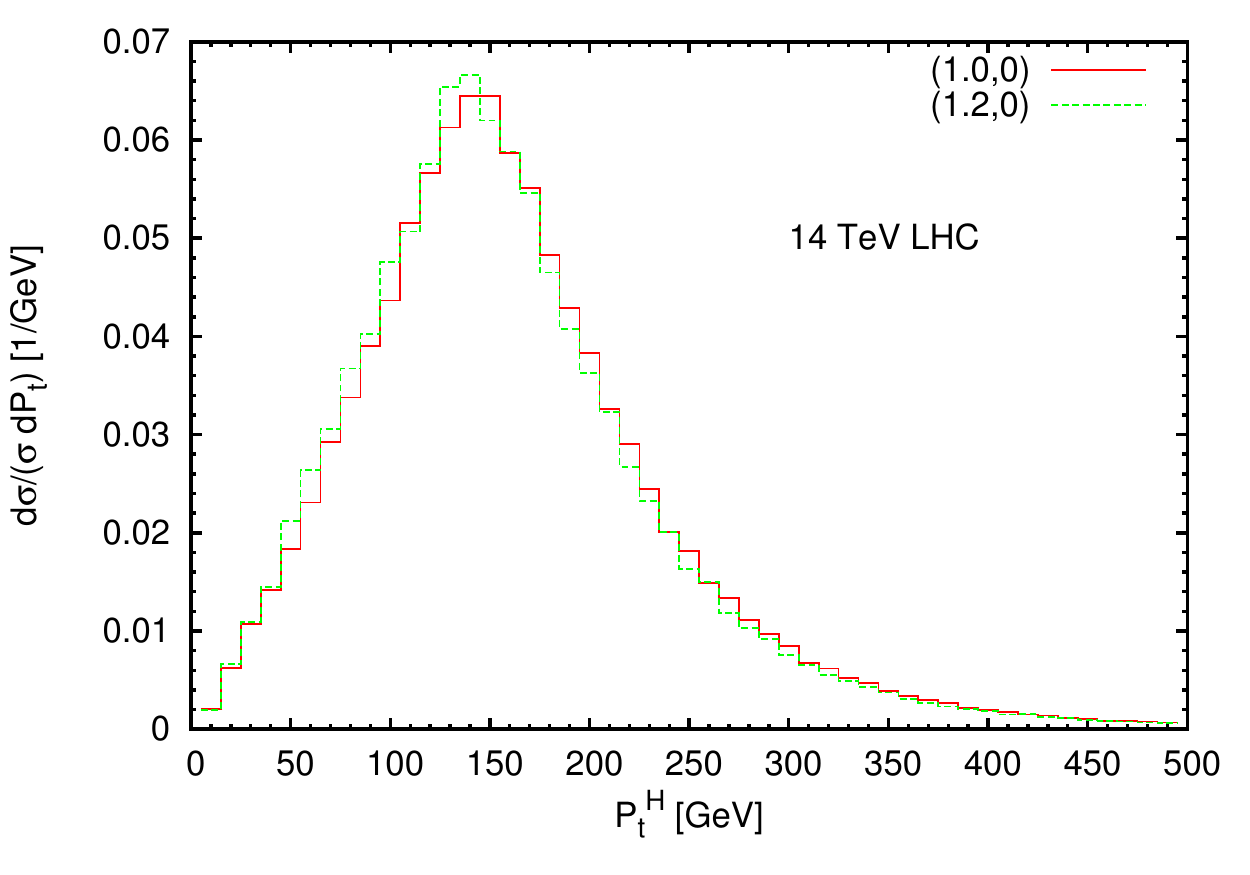}
  \captionsetup{width=.8\textwidth}
   \caption{Comparison of normalized $P_t$-distributions of the Higgs for $a=1.2, b=0$ case and the standard model case.}
  \label{fig:ptplotbm1}
\end{minipage}%
\begin{minipage}{.5\textwidth}
  \centering
  \includegraphics[width=7.5cm]{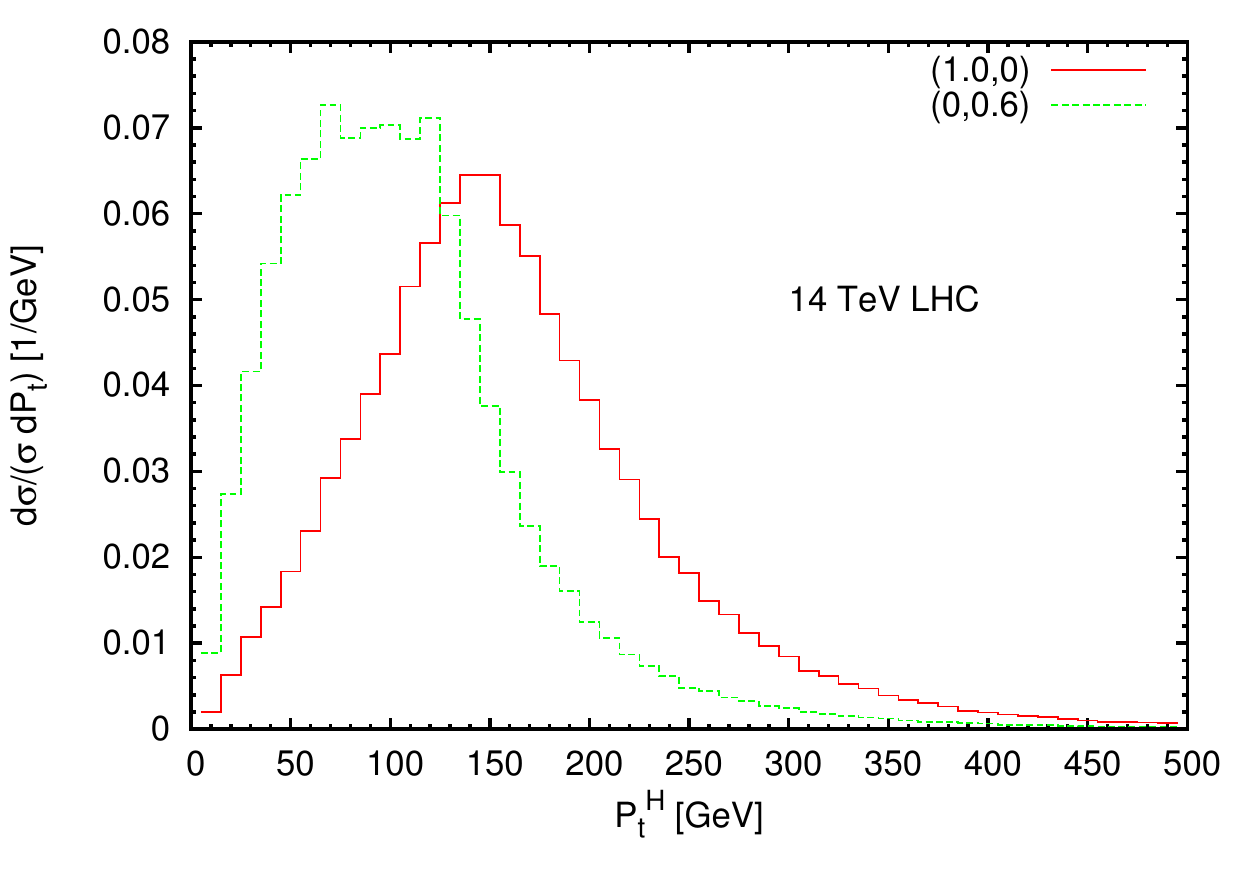}
  \captionsetup{width=.8\textwidth}
  \caption{Comparison of normalized $P_t$-distributions of the Higgs for $a=0, b=0.6$ case and the standard model case.}
  \label{fig:ptplotbm2}
\end{minipage}
\end{figure}
\begin{figure}[ht!]
\centering
\begin{minipage}{.5\textwidth}
  \centering
  \includegraphics[width=7.5cm]{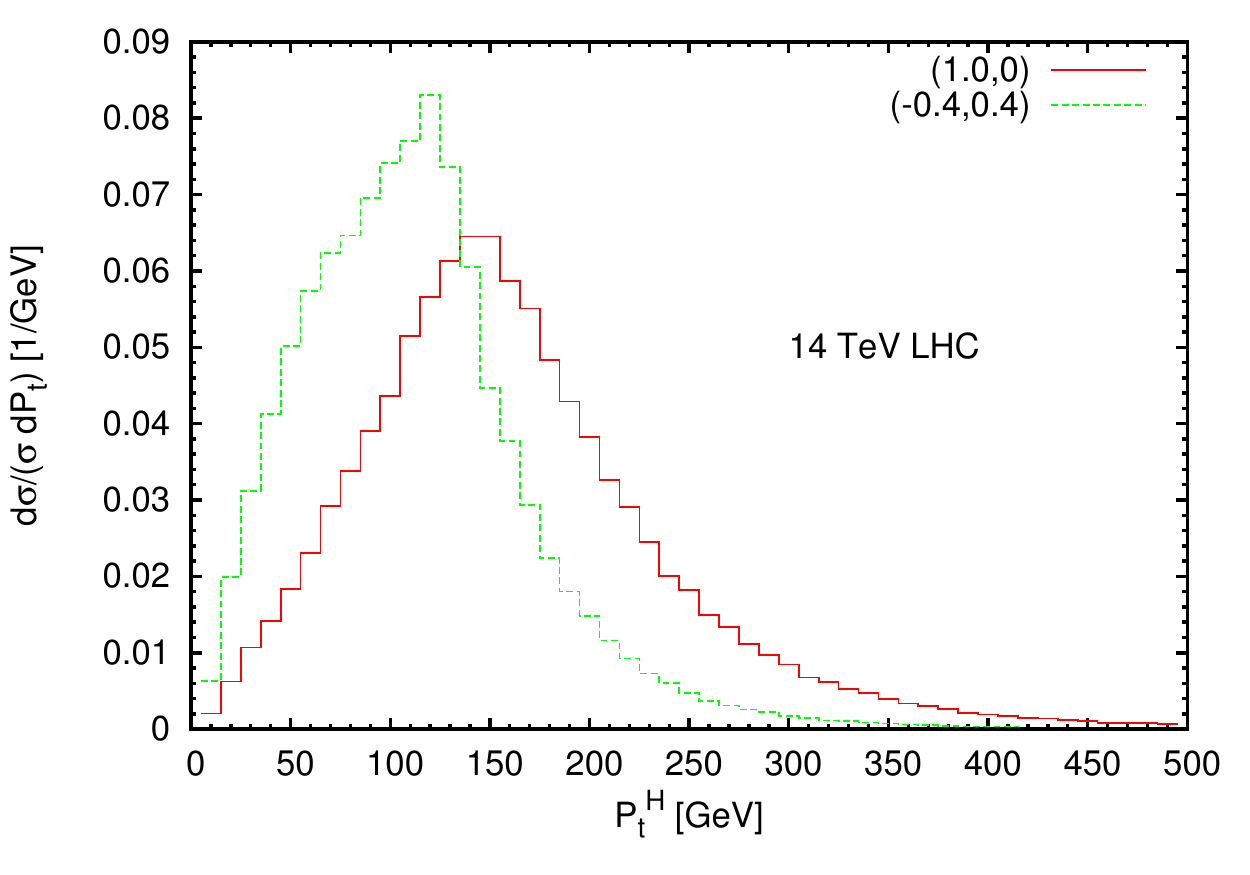}
  \captionsetup{width=.8\textwidth}
   \caption{Comparison of normalized $P_t$-distributions of the Higgs for $a=-0.4, b=0.4$ case and the standard model case.}
  \label{fig:ptplotbm3}
\end{minipage}%
\begin{minipage}{.5\textwidth}
  \centering
  \includegraphics[width=7.5cm]{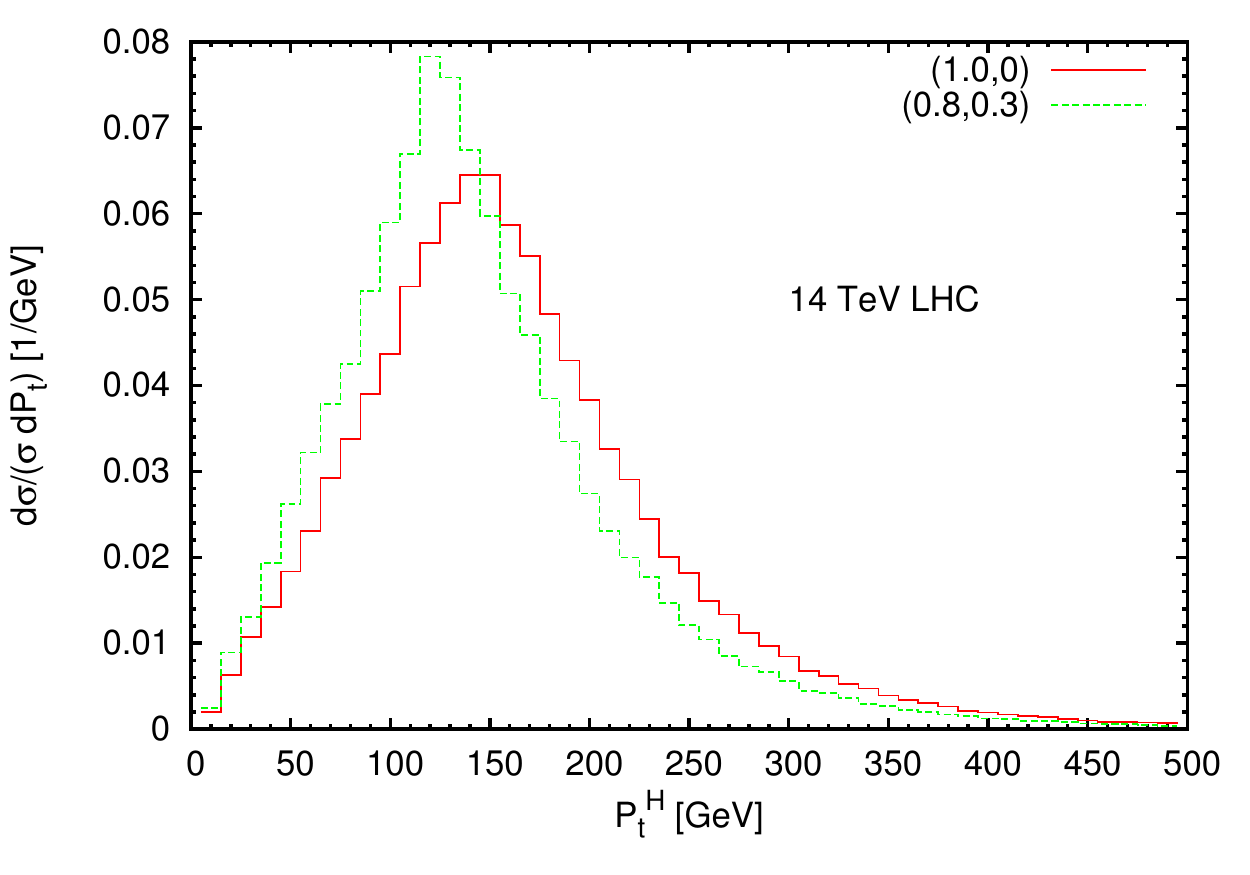}
  \captionsetup{width=.8\textwidth}
  \caption{Comparison of normalized $P_t$-distributions of the Higgs for the best-fit values $a=0.8, b=0.3$ case and the standard model case.}
  \label{fig:ptplotbm4}
\end{minipage}
\end{figure}

 Now and here, we again have a discussion on the double Higgs production after choosing four benchmark values of 
the parameters $(a,b)$ which are allowed by the present LHC data. 
For these benchmark values, the two Higgs production cross sections at 8, 14 and 33 TeV center-of-mass energies 
are given in table 3. The table suggests that the cross section might get enhanced or reduced within the allowed 
parameter space. 
The Higgs $P_t$ distributions in all the four cases are compared with the standard model case in 
Figs.~\ref{fig:ptplotbm1}-\ref{fig:ptplotbm4}. We have plotted them 
separately to emphasize the deviations in each case. These are consistent with the observations made in Sec. 2. 
Like the deviations in $P_t$ distributions, the deviations in 
invariant mass distributions $M_{HH}$ are also not very large for $(1.2,0)$ and $(0.8,0.3)$ cases.

\begin{figure}[t]
\centering
\begin{minipage}{.5\textwidth}
  \centering
  \includegraphics[width=7.5cm]{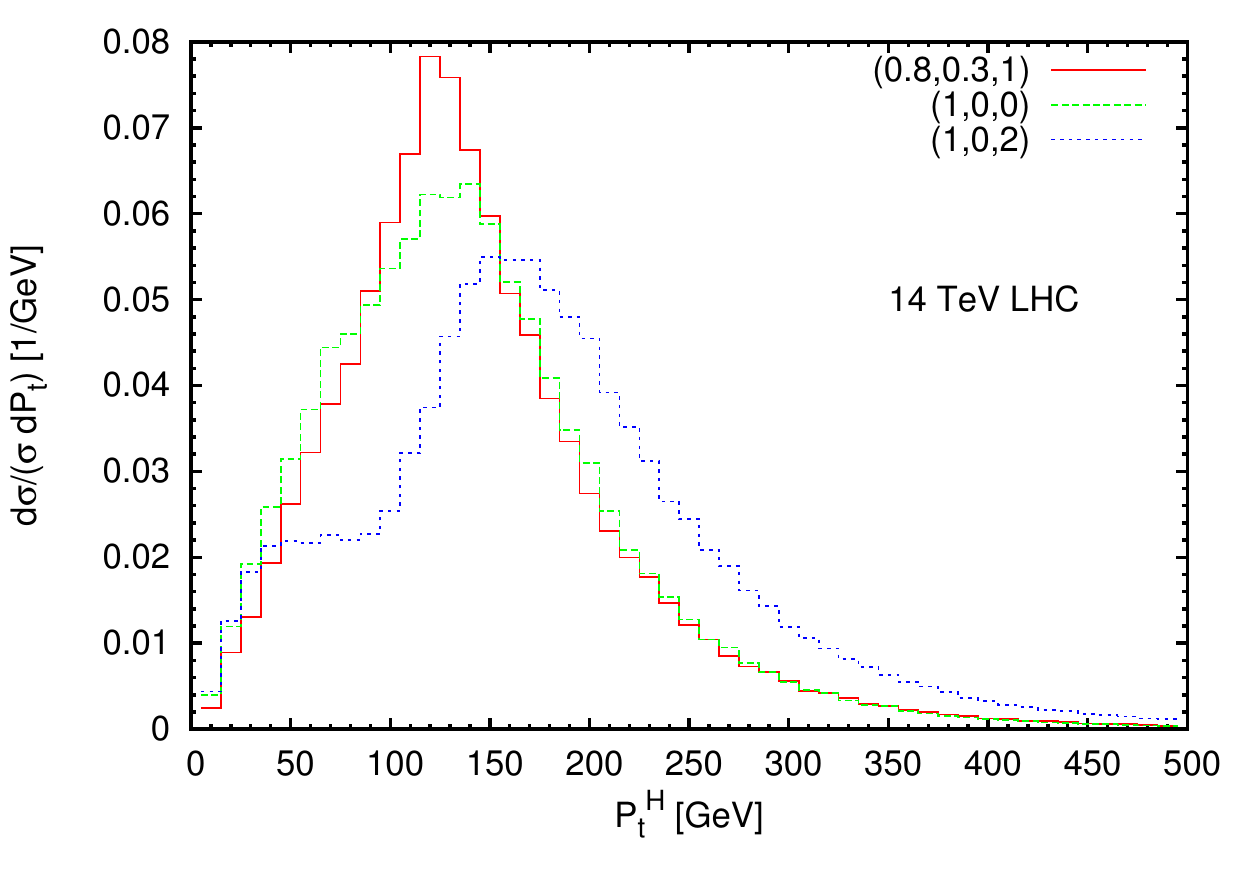}
  \captionsetup{width=.8\textwidth}
   \caption{Normalized $P_t$-distributions of the Higgs for various combinations of parameters $(a,b,\kappa)$.}
  \label{fig:ptplot2}
\end{minipage}%
\begin{minipage}{.5\textwidth}
  \centering
  \includegraphics[width=7.5cm]{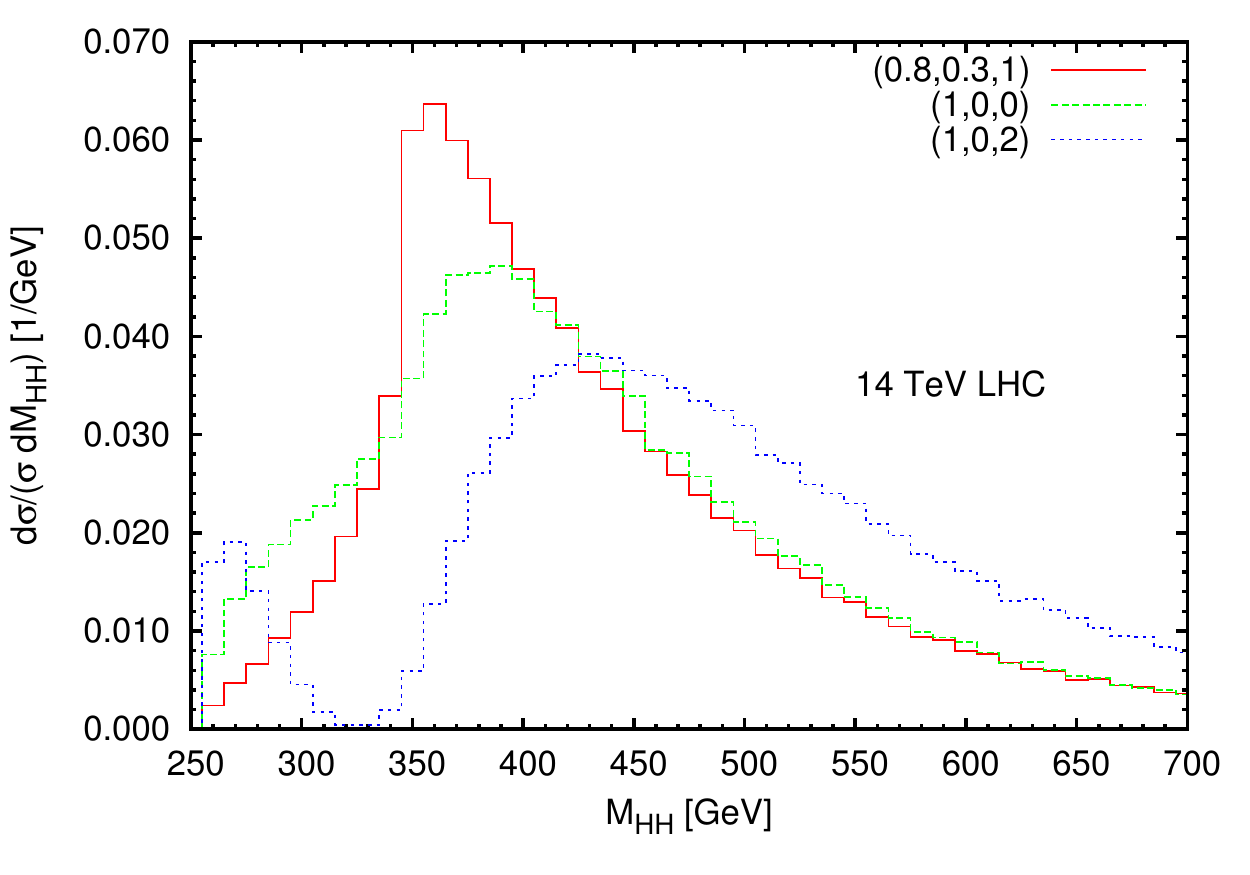}
  \captionsetup{width=.8\textwidth}
  \caption{Normalized two Higgs invariant mass distributions for various combinations 
          of $(a,b,\kappa)$.}
  \label{fig:mhhplot2}
\end{minipage}
\end{figure}
The double Higgs production process has also been studied in the context of anomalous 
trilinear Higgs coupling ($\lambda_{HHH}$). It is therefore important to investigate if there may be any overlap between
the predictions due to the $ttH$ anomalous coupling and those due to the anomalous trilinear
Higgs coupling. We define the anomalous trilinear Higgs coupling using, $\lambda_{HHH} = \kappa \lambda_{HHH}^{\rm SM}$, 
where $\lambda_{HHH}^{\rm SM}$ is the standard model value of the coupling. 
Here we take, $\kappa=0,1,2$ as possible values of the scale factor, $\kappa =1$ being the standard model case.
{The Higgs pair production rates in presence of the anomalous trilinear Higgs coupling, are added
in the last two columns of table 3. Note that in the case of $\kappa=0$ only the box amplitude contributes to the 
cross section. In $\kappa=2$ case, the triangle contribution increases and the destructive 
interference between box and triangle amplitudes becomes more severe. The enhanced destructive interference
effect, in this case, is visible in the kinematic distributions, shown in Figs.~\ref{fig:ptplot2} 
and \ref{fig:mhhplot2}. In these figures, the kinematic distributions for $\kappa=0$ and $\kappa=2$ cases 
are compared with those for the best-fit values of $(a,b)$.
Due to characteristic differences in the distributions and very different values of 
cross sections, it should be possible to discriminate the case of the anomalous trilinear Higgs coupling from the 
case of the anomalous top-Higgs coupling. The possibility of the introduction of both the anomalous couplings 
may lead to more interesting situations.\footnote{
 {In pure pseudoscalar $ttH$ coupling case, the unpolarized 
cross section does not depend on the scale factor $\kappa$.}}}


 \begin{figure}[t]
 \includegraphics[width=3.15in]{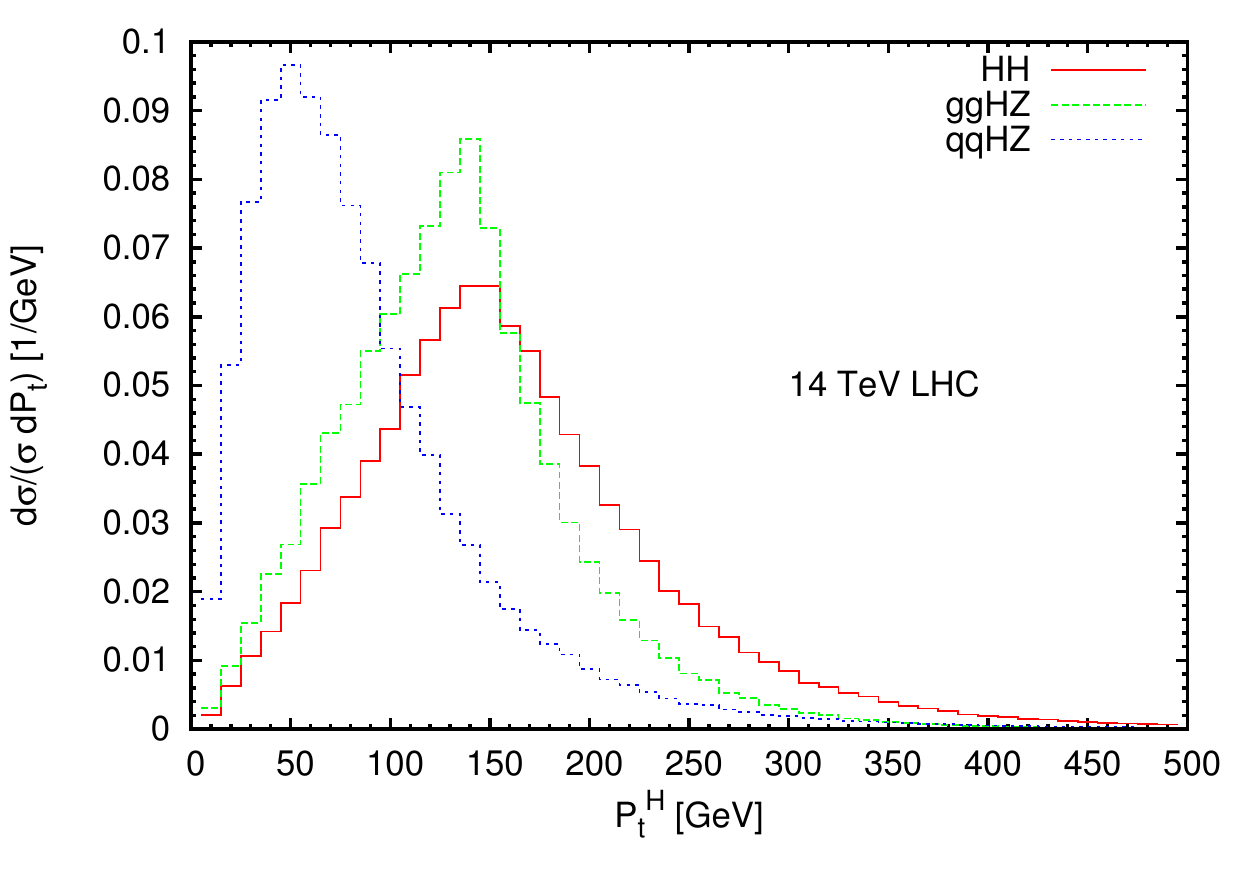}
 \includegraphics[width=3.15in]{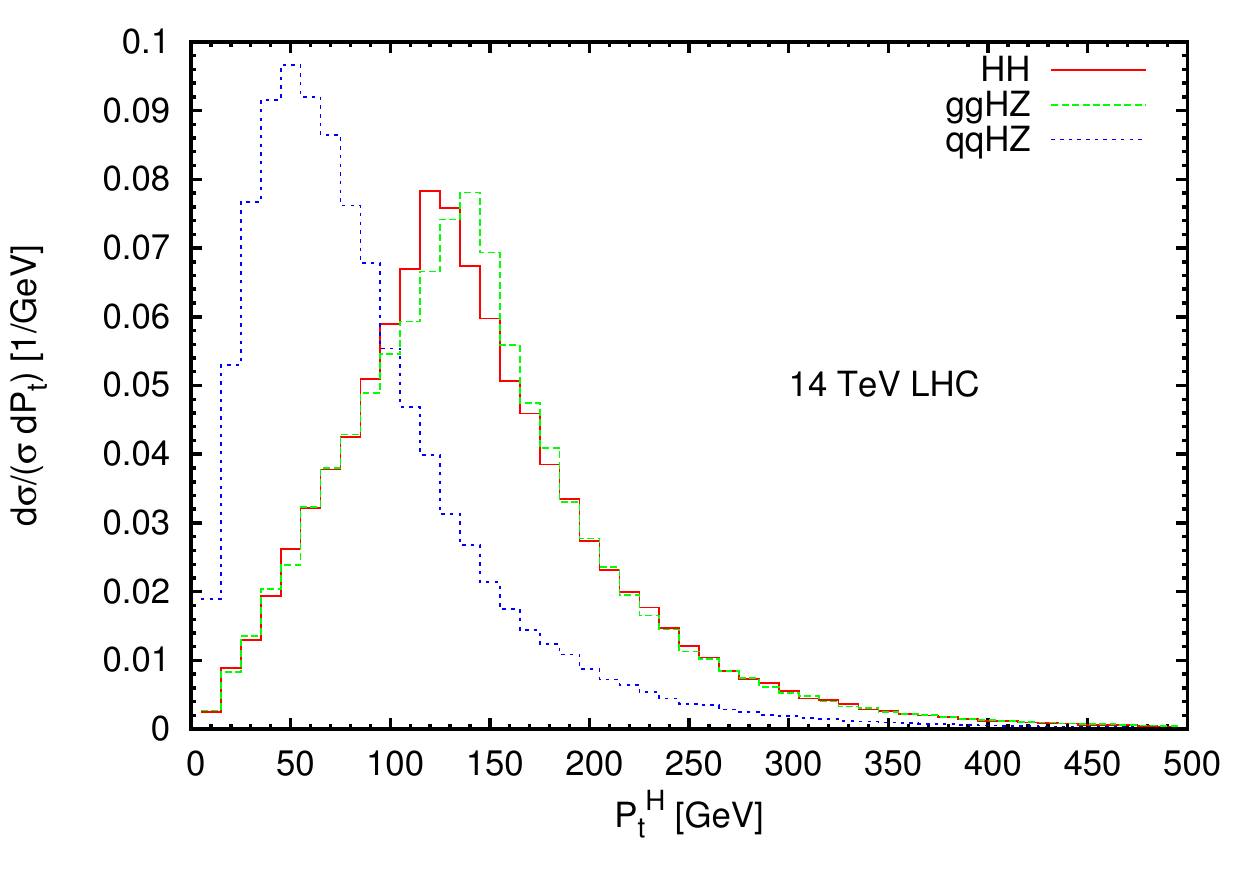}
 \caption{Comaprison of normalized $P_t$ distributions of the Higgs pair production and $HZ$ background processes. 
          The contribution from quark-quark channel to $pp \to HZ$ is calculated at the tree level. The plot on the left
          is the standard model case, while the right one corresponds to the best-fit values $a=0.8, b=0.3$.}
 \label{fig:ptplotHZ}
 \end{figure}

Out of many decay channels, the $HH \to b\bar{b}\gamma\gamma$ is the most promising channel to observe 
double Higgs production at the LHC. As described in the {Ref.}~\cite{Baglio:2012np}, $ZH$ production process
is one of the main backgrounds in this channel.
In the standard model, the tree-level cross section for $pp \to ZH$ {at $14\,\text{TeV}$} is about
$0.63$ pb and the $K$-factor at the next-to-next-to-leading order (NNLO) in QCD is close to 1.33~\cite{Dittmaier:2011ti}. 
A part of the NNLO QCD contribution which arise 
due to the gluon-gluon fusion is {also important} at the LHC. Its cross section is $\sim$ 100 fb at 14 
TeV. In  Fig.~\ref{fig:ptplotHZ}, we can see the relative importance of the gluon-gluon channel over the 
quark-quark channel in higher $P_t$ region. Note that these distributions are normalized. 
Due to the much larger quark-quark channel contribution, the peak of the combined distribution does not 
shift from its tree-level position. A large $P_t$ cut 
can be applied to suppress the contribution coming from the quark-quark channel.
We also notice a significant overlap of 
the Higgs $P_t$ distributions in $gg \to HH$ and $gg \to ZH$ cases in the standard model. The Higgs $P_t$ distributions 
are also compared for the best-fit values of the parameters $a$ and $b$ in Fig.~\ref{fig:ptplotHZ}.

 \begin{table}[t]
 \begin{center}
  \begin{tabular}{|c|c|c|c|c|c|c|}
   \hline
   $\sqrt{\rm S}$& $\sigma_{(1,0)}$& $\sigma_{(1.2,0)}$& $\sigma_{(0,\pm0.6)}$& $\sigma_{(-0.4,\pm0.4)}$&  $\sigma_{(0.8,\pm0.3)}$ \\
   (TeV) & (fb)& (fb)& (fb)& (fb) & (fb)\\
   \hline
    8 &  24.72 &  20.35 &   63.58  &   80.96  &  31.01     \\
   \hline
   14 &  97.98 &  79.42 &  275.36  &  355.74  & 126.08     \\
   \hline
   33 & 569.60 & 454.15 & 1788.12  & 2346.59  & 756.69     \\
   \hline
  \end{tabular} 
 \end{center}
 \caption{$gg \to ZH$ hadronic cross sections for allowed benchmark values of parameters $(a,b)$. The kinematic 
          settings in this case are same as in the two Higgs production case.}
\end{table}

Diagrams contributing to $gg\to ZH$ amplitude are quite similar 
to the case of double Higgs production, however, only box diagram involves the top-Yukawa coupling. The gluon-gluon 
channel to $ZH$ production thus becomes very important background {for the Higgs pair production process in presence of}
anomalous $ttH$ coupling. The effect of anomalous top-Higgs coupling on the $gg \to ZH$ cross section 
at various collider center-of-mass energies are listed in table 4. The contributions from both the 
top and bottom quarks are included to cancel the anomaly in triangle diagram. Like the two Higgs production 
case, the box and triangle amplitudes interfere destructively in $ZH$ case. The triangle amplitude, however, 
dominates the cross section. Due to this the cross section for the $(1.2,0)$ case is smaller than the standard 
model cross section. We also note that non-zero $b$ can introduce large enhancement in the cross section. 
In fact, the $gg \to HZ$ channel can be separately studied to probe the anomalous top-Higgs coupling at the LHC.

We have already seen that due to top-Higgs anomalous coupling, the $P_t^H$ and $M_{HH}$ distributions in the two Higgs 
production case
shift towards low transverse momentum and low invariant mass regions. 
Referring back to the signal-background analysis performed in Ref.~\cite{Baglio:2012np} in $b\bar{b}\gamma\gamma$ 
channel, we note that the suggested cuts on $P_t^H$ and  $M_{HH}$ may, therefore, not be effective in presence of anomalous 
$ttH$ coupling. Nevertheless, cuts on $\eta_H$ and $\eta_{HH}$ may still be useful. Probing the 
effects of anomalous $ttH$ coupling in the two Higgs production process at the LHC turns out to be a challenging task.  
It is clear that if we observe higher rates for the Higgs pair production at the LHC, it may not 
be only due to the top-Higgs anomalous coupling under consideration. It should be noted that large enhancement in the 
cross section can be realised only in some limited parameter space. Therefore, if lower production rates are observed 
this coupling can provide an explanation. 
This will require a more complete and detailed collider study which is beyond the scope of the present work.

\section*{Acknowledgement}
We thank Shankha Banerjee and {Anushree Ghosh} for useful discussions and comments on global analysis.
We would also like to thank Adam Falkowski, Ushoshi Maitra, Eduard Masso, Biswarup Mukhopadhyaya
and Santosh Kumar Rai for their remarks on various issues 
related to the present work.
The authors are partially supported by funding available from the Department of 
Atomic Energy, Government of India for the Regional Centre for Accelerator-based
Particle Physics (RECAPP), Harish-Chandra Research Institute.

\end{document}